\documentclass[fleqn,10pt]{wlscirep}

\usepackage[utf8]{inputenc}
\usepackage[T1]{fontenc}
\usepackage{tikz}
\usepackage{lscape}
\usetikzlibrary{shapes, arrows,positioning}
\usetikzlibrary{matrix,shapes,arrows,positioning}

\usetikzlibrary{shadows,arrows.meta,positioning,backgrounds,fit}

\usepackage{amssymb,amsfonts,latexsym,amsmath,amsthm,times}
\usepackage{graphicx}
\usepackage{epstopdf}
\usepackage{tikz-qtree}

\graphicspath{ {./images/} }
\usepackage[rightcaption]{sidecap}

\usepackage{wrapfig}
\usepackage{color}
\usepackage[Conny]{fncychap}
\usepackage{enumitem}

\usepackage{algorithmic}
\usepackage{float}

\usepackage{lineno,hyperref}
\usepackage[edges]{forest}
\usetikzlibrary{arrows.meta}
\usepackage{pdflscape}
\usepackage{pgfkeys}
\usepackage{caption}
\usepackage{multirow}
\usepackage{multicol}
\usepackage{bigstrut}
\usepackage{array}
\usepackage{colortbl}

\usepackage{arydshln}
\usepackage{cleveref}
\usepackage{setspace} 
\usepackage{pdfpages}

\usepackage{xr}
\usetikzlibrary{calc,trees,positioning,arrows,chains,shapes.geometric,%
    decorations.pathreplacing,decorations.pathmorphing,shapes,%
    matrix,quotes,shapes.symbols,plotmarks}
    \usepackage[paper=portrait,pagesize]{typearea}

    \tikzset{parent/.style={align=center,text width=2cm,,rounded corners=2pt},
child/.style={align=center,text width=2.8cm,rounded corners=6pt},
grandchild/.style={text width=2.3cm}
}

\forestset{%
  dir tree switch forking/.style args={at #1}{%
    for tree={
      font=\sffamily,
      fit=rectangle,
    },
    where level=#1{
      for tree={
        folder,
        grow'=0,
      },
      delay={
        child anchor=north,
        !u.parent anchor=south,
        edge path'={(!u.parent anchor) -- ++(0,-\forestoption{fork sep}) -| (.child anchor)}
      },
    }{
      if={>On<{level}{#1}}{
        forked edge,
        parent anchor=children,
        child anchor=parent,
      }{},
    },
    before typesetting nodes={
      for tree={
        if content={}{
          coordinate,
          no edge,
          before packing={!u.l sep'=0pt},
          before computing xy={s/.option=!u.min x}
        }{
          content/.wrap value={\strut ##1},
        },
      },
      for nodewalk={filter={tree}{>On<{level}{#1}}}{align to centre},
    },
  },
  align to centre/.style={
    if={>Ow+P{n children}{isodd(#1)}}{
      for nodewalk/.process={Ow+nw{!r.n children}{(#1+1)/2}{{fake=r,n=#1}{calign with current edge}}},
    }{
      calign=edge midpoint,
    },
  },
}



\newcolumntype{L}[1]{>{\centering\arraybackslash}m{#1}}

\title{Insecticide-treated bed net use and elimination of malaria in Sub-Saharan African countries: Assessing the Global Technical Strategy using an evolutionary game approach}

\author[1]{Laxmi}
\author[2]{Tamer Oraby}
\author[3]{Michael G Tyshenko}
\author[4]{Ina Danquah}
\author[1,*]{Samit Bhattacharyya}
\affil[1]{Disease Modelling Lab, Department of Mathematics, School of Natural Sciences, Shiv Nadar Institution of Eminence, India}
\affil[2]{School of Mathematical and Statistical Sciences, The University of Texas
Rio Grande Valley, Edinburg, TX 78539, USA
}
\affil[3]{Risk Sciences International, Ottawa, ON, K1P 5J6, Canada}

\affil[4]{Center for Development Research (ZEF), University of Bonn
Genscherallee 3, 53113 Bonn, Germany}

\affil[*]{Corresponding author: samit.b@snu.edu.in }

\begin{abstract}
Malaria continues to be a major public health challenge in Sub-Saharan Africa (SSA), where most countries have failed to meet the World Health Assembly’s endorsed Global Technical Strategy (GTS) milestones for malaria reduction. While insecticide-treated net (ITN) usage is a proven intervention, many survey studies suggest that improper net use is likely a significant barrier to success. However, it remains an untested hypothesis. For the first time, we present the application of a behaviour-incidence model to inform ITN use and GTS outcomes. This model of ITN use was fitted to malaria case data to better explain SSA countries’ varying progress toward GTS targets. Unlike previous studies focusing on ITN coverage and efficacy, we categorized the 38 SSA countries into achievers, non-achievers, and outliers based on model outcomes. Results indicate that while some countries can meet GTS 2025 and 2030 targets with ongoing efforts, others require enhanced social awareness campaigns, economic assistance, and improvements in ITN efficacy to succeed. Our analysis emphasizes country-specific behavioural interventions are essential for accelerating malaria elimination. Finally, our model provides actionable policy insights, emphasizing tailored strategies to optimize proper ITN use, which will ensure all SSA countries can achieve GTS malaria elimination targets.

\vspace{0.25cm}


\vspace{0.25cm}
\textbf{Word count:} 5849
\end{abstract}

\usepackage{xr}
\makeatletter

\newcommand*{\addFileDependency}[1]{
\typeout{(#1)}
\@addtofilelist{#1}
\IfFileExists{#1}{}{\typeout{No file #1.}}
}\makeatother

\newcommand*{\myexternaldocument}[1]{%
\externaldocument{#1}%
\addFileDependency{#1.tex}%
\addFileDependency{#1.aux}%
}

\myexternaldocument{supplementary}

\begin{document}

\pagestyle{plain}
\flushbottom
\maketitle
%
%
\thispagestyle{empty}

\doublespacing
\section*{Introduction}
Malaria is an infectious disease transmitted to humans primarily through the bite of malaria-infected female \textit{Anopheles} mosquitoes \cite{savi2022overview, amambua2019major}. The highest incidence and mortality rates are reported in the African region alone \cite{world2022world}. Twenty-nine countries accounted for $96\%$ of malaria cases globally, and six countries—Nigeria ($27\%$), the Democratic Republic of the Congo ($12\%$), Uganda ($5\%$), Mozambique ($4\%$), Angola ($3.4\%$) and Burkina Faso ($3.4\%$)- accounted for about $55\%$ of all cases globally\cite{world2023world, oladipo2022increasing}. Insecticide-treated nets (ITNs) have proven to be one of the most widespread, effective, and cost-efficient tools for combating malaria in Africa, averting an estimated 450 million cases from 2000-2015 \cite{bhatt2015effect, bertozzi2021maps}.\\
Despite a notable expansion of malaria interventions, including the distribution of ITNs and other measures, global malaria mortality rates declined by $60\%$ between 2001 and 2015. However, the disease remains endemic, with the highest burden in the African region, where an estimated 90\% of all malaria deaths occur \cite{world2015achieving, world2015global}. Although the implementation of core interventions expanded greatly between 2000 and 2014, the gains are fragile and unevenly distributed. In pursuit of reducing the burden of disease and eliminating malaria, the Global Technical Strategy (GTS) was adopted by the 68th World Health Assembly in May 2015 \cite{world2015global}. This technical strategy provides a framework for the development of tailored programs to accelerate progress toward malaria elimination \cite{WHO2018, kearney2024geospatial, newby2016path}. It defines a clear and ambitious path for countries in which malaria is endemic and their global partners in malaria control and elimination for the next 15 years. As part of the vision of the WHO and the global malaria community, making a world free of malaria, the strategy sets goals to achieve global targets for 2030 with milestones to measure progress for 2020 and 2025 as mentioned in Table\ref{GTS}.

\begin{table}
\centering
\caption{\textbf{Vision: A World Free of malaria:} Goals, Milestones and Targets for the Global Technical Strategy for Malaria 2016-2030.\cite{world2015global}}
\resizebox{\textwidth}{!}
{\begin{tabular}{>{\columncolor{gray!40}}L{5cm}>{\columncolor{gray!10}}L{3cm}>{\columncolor{gray!10}}L{3cm} >{\columncolor{gray!40}}L{3cm}}
\multicolumn{4}{l}{\Large \textbf{}}\\[2pt]
\hline
\textbf{Goals} & \multicolumn{2}{c}{\textbf{Milestones}} & {\textbf{Targets}} \\[1pt]
  \hline
   & 2020 & 2025 & 2030\\
   \cdashline{2-4}
1. Reduce malaria mortality rates globally compared with 2015 & At least 40$\%$ & At least 75$\%$ & At least 90$\%$ \\
2. Reduce malaria case incidence globally compared with 2015 & At least 40$\%$ & At least 75$\%$ & At least 90$\%$ \\
3. Eliminate malaria from countries in which malaria was transmitted in 2015 & At least 10 countries & At least 20 countries & At least 35 countries \\ 
4. Prevent re-establishment of malaria in all countries that are malaria-free & Re-establishment prevented & Re-establishment prevented & Re-establishment prevented \\ \hline

\end{tabular}}
\label{GTS}
\end{table}

\begin{figure}
     \centering
    \includegraphics[width =1.2\textwidth]{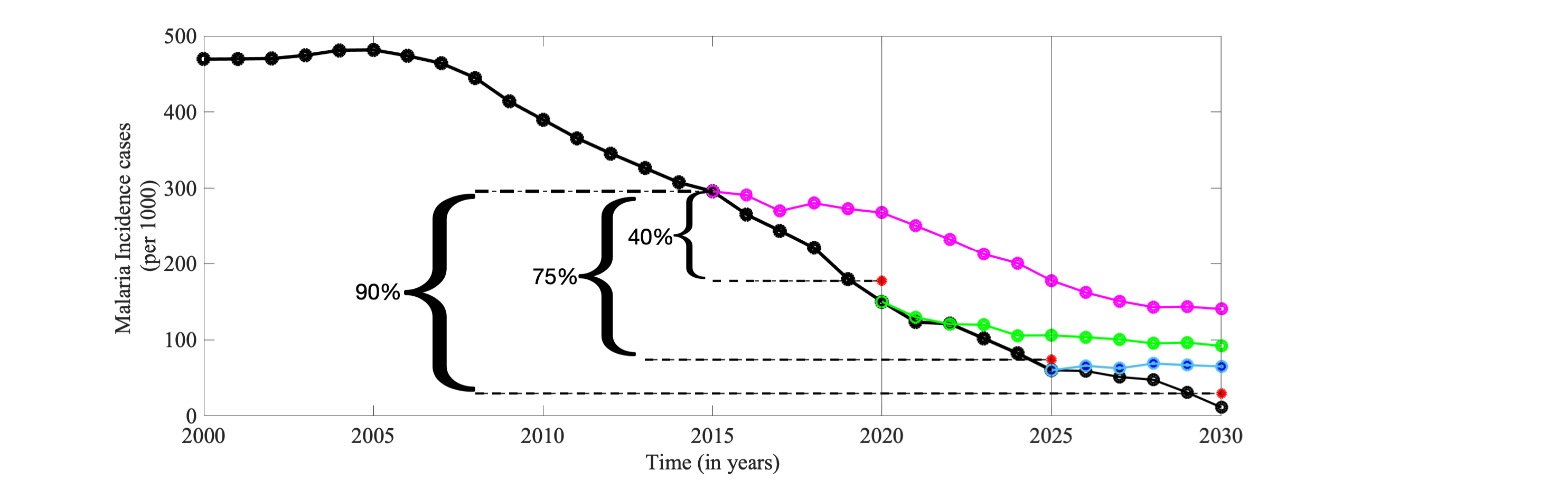}
    \caption{An illustration of Global Technical Strategy (GTS) milestones over time, based on malaria incidence is presented. Using 2015 as the baseline, the 2020 milestone is defined as a 40\% reduction in incidence by 2020, followed by a 75\% reduction by 2025, and a 90\% reduction by 2030. The curve marked with black dots represents an incidence trajectory that meets all milestones. The pink curve fails to achieve the milestones from 2020 onwards, the green curve from 2025 onwards, and the blue curve misses only the 2030 milestone. The period from 2000 to 2020 is referred to as \textbf{Phase-0}, 2020 to 2025 as \textbf{Phase-I}, and the next phase up to 2030 is \textbf{Phase-II}.}
    \label{fig:ideal GTS}
\end{figure}

\begin{figure}
     \centering
     \includegraphics[width =1\textwidth]{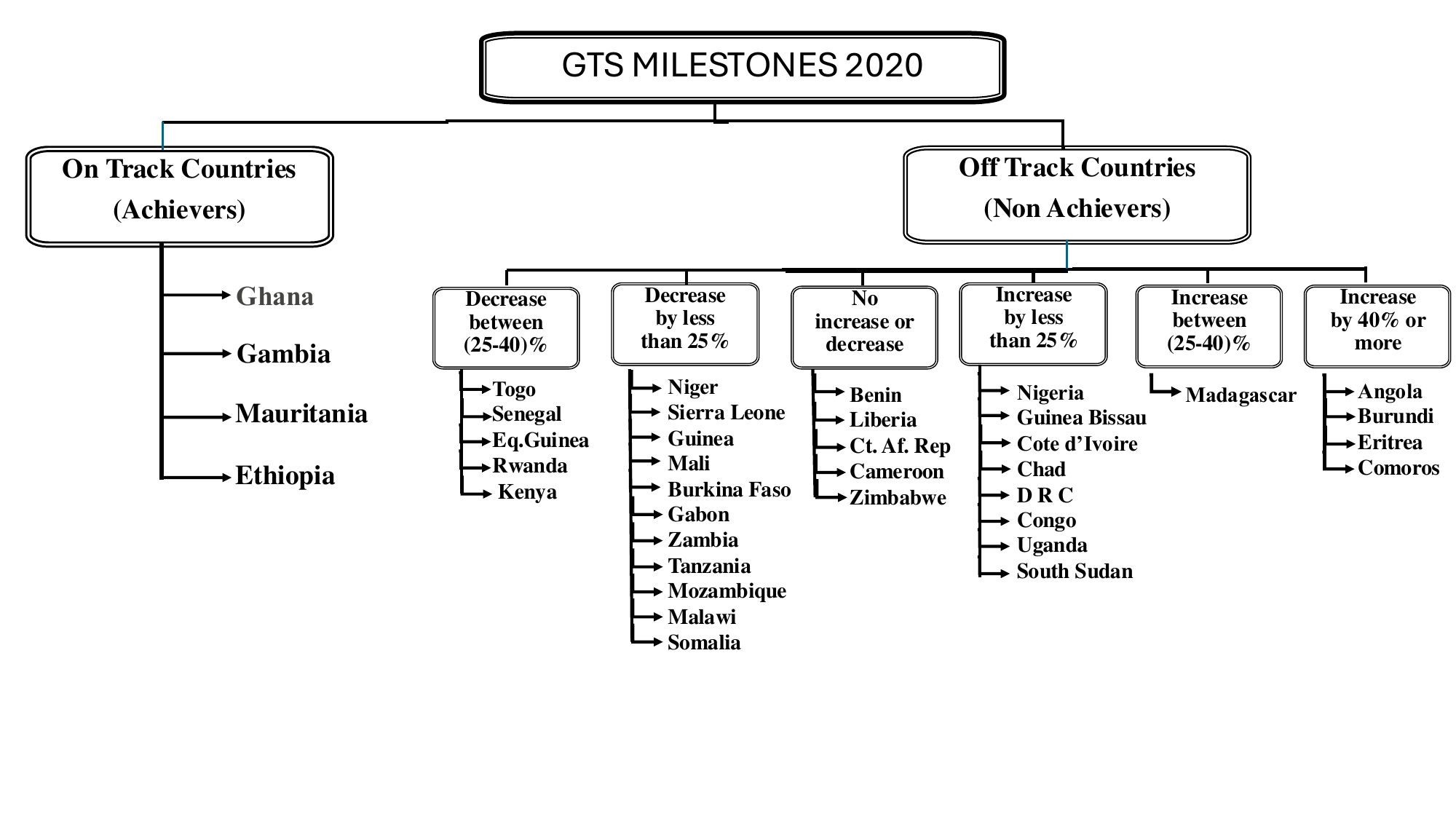}
     \vspace{-1cm}
     \caption{Classification of 38 Sub-Saharan African (SSA) countries as either achievers or non-achievers of the 2020 Global Technical Strategy (GTS) milestones.}
\label{fig:Decision Tree}
 \end{figure}

Figure \ref{fig:ideal GTS} provides a schematic representation of the ideal trajectory towards meeting GTS goals, along with some alternative trajectories that highlight potential deviations from the optimal path. 

But WHO malaria report 2021 \cite{world2022world} revealed that GTS 2020 milestones for morbidity and mortality, based on the 2015 baseline, have not been achieved. Globally, the world is off track by 42$\%$ and if this trajectory continues, by 2030 it will be off track by 91$\%$. The report presents an analysis of the trends by region that shows WHO African Region is off track for both malaria morbidity and mortality GTS 2020 milestones, by $38\%$ and $40\%$ respectively. Despite the extensive distribution of insecticide-treated nets (ITNs), one of the most widespread, cost-effective, and impactful interventions in Sub-Saharan Africa, the trends in estimated malaria cases and deaths remain concerning, posing serious challenges to achieving the milestones and goals of the Global Technical Strategy (GTS) for 2025 and 2030. Only 4 of the 40 countries with the highest burden in SSA met GTS 2020. The list of achievers (on track) and non-achievers (off-track) SSA countries in different subcategories is given in Figure \ref{fig:Decision Tree}.\\

The failure to meet the GTS 2020 milestones in Sub-Saharan Africa stems from a complex interplay of socioeconomic, biological, and behavioural factors that hinder malaria control efforts \cite{world2015global}. Unlike other regions where malaria control has advanced more rapidly, Sub-Saharan Africa faces unique challenges \cite{okumu2022africa}. For example, biological threats such as widespread resistance to insecticides and antimalarial drugs undermine the effectiveness of current control strategies in SSA countries \cite{world2022world}. In addition, high levels of poverty in the region have historically hampered efforts to combat infectious diseases such as malaria. Poverty not only limits access to healthcare and preventive measures such as insecticide-treated nets (ITNs), but it also affects people's ability to replace or maintain these nets over time. Furthermore, poverty is often associated with limited education and low awareness, leading to misuse of ITNs for non-health-related purposes such as fishing, fencing, and protecting seedlings. These widespread behavioural practices in Sub-Saharan African countries considerably hinder the progress toward the goals of malaria elimination. \cite{asingizwe2019role, okumu2022africa}.\\

\begin{figure}[h]
     \hspace{0cm}
    \includegraphics[width =1\textwidth]{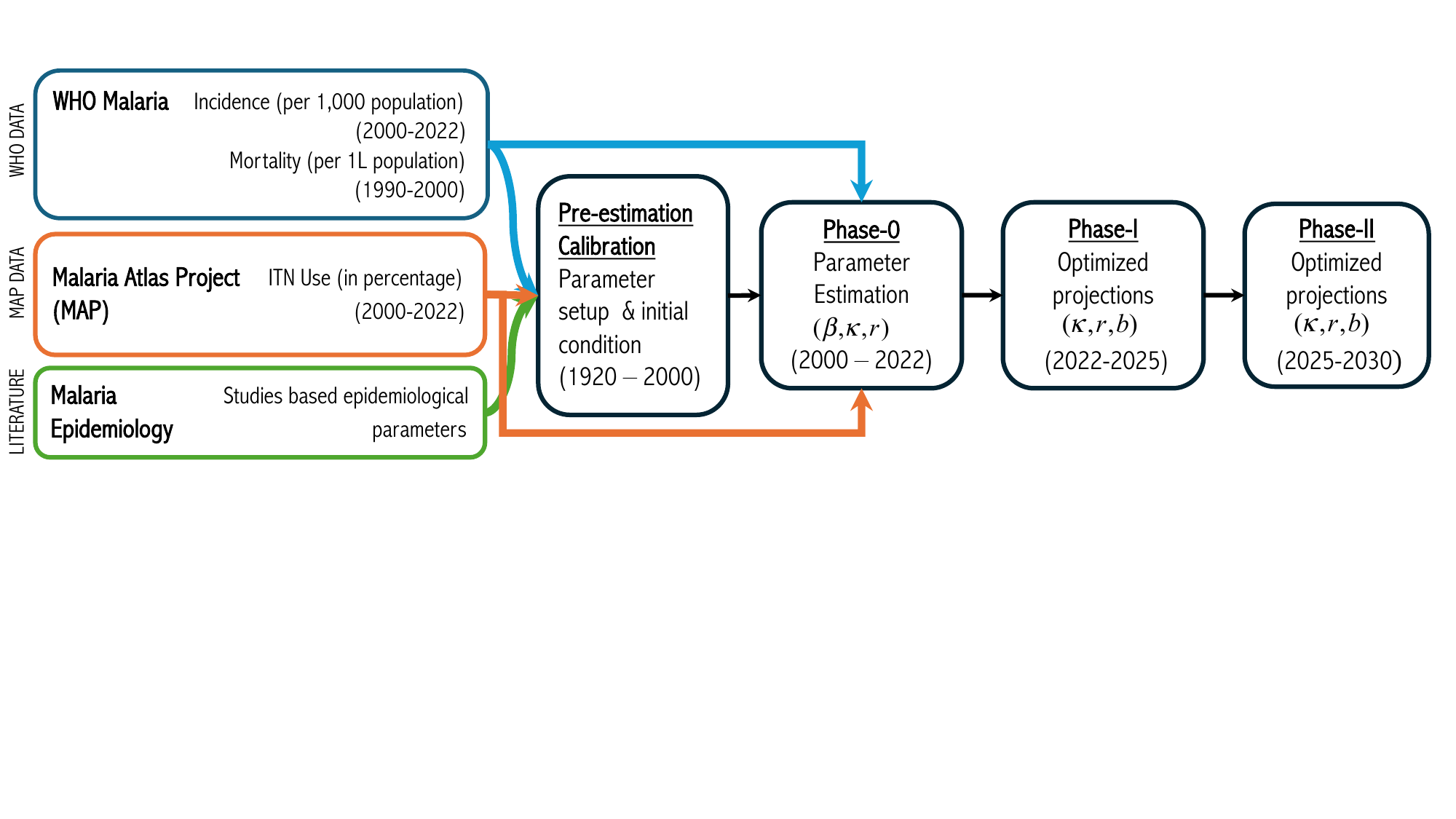}
    \vspace{-3cm}
    \caption{Analysis Workflow Schematic: WHO and MAP data, along with literature-based parameter values for malaria, were utilized for pre-estimation calibration of the behaviour-incidence model. These data were further employed to estimate behavioural and transmission parameters. The calibrated model was subsequently optimized and projected to assess malaria incidence and ITN use, enabling evaluation and prediction of country-specific GTS targets.}
    \label{fig:flow_diagram}
\end{figure}

There are numerous ITN-related research studies, including modelling malaria \cite{mandal2011mathematical}, that address various aspects of malaria prevention and control. For instance, many studies explore the efficacy of ITNs in malaria control \cite{lindsay2021threats}, the spatiotemporal coverage of ITNs across different SSA countries \cite{bertozzi2021maps}, the relationship between ITNs and insecticide resistance \cite{oxborough2024malaria}, and the role of women's empowerment in ITN usage for malaria prevention \cite{kwansa2024women, kanmiki2019socio}. Other research focuses on quantifying the direct and indirect protection provided by ITNs against malaria \cite{unwin2023quantifying}, social awareness and caregivers' knowledge about ITN use among children under five \cite{kumah2024influence}, and factors influencing ITN usage among pregnant women in Ghana \cite{alhassan2022impact}, among others. Despite the abundance of studies on ITN efficacy, usage, and their impact on malaria control in SSA countries, there remains a notable gap in research on human behavioural practices related to ITN use and their implications for malaria elimination, particularly within the framework of the Global Technical Strategy. Such analyses are vital for developing a deeper understanding and optimizing malaria control measures tailored to specific countries, thereby enhancing localized disease management and malaria elimination policies.\\

In this research, we performed the analysis in multiple steps (Figure \ref{fig:flow_diagram}). Firstly, we explored the dynamic association between malaria prevalence and ITN usage in Sub-Saharan Africa using transfer entropy and other statistical methods. Secondly, we used the integrated behaviour-incidence model, performed parameter estimation technique and optimization, and leveraged empirical data on malaria incidence and use of ITN to identify key behavioural and epidemiological parameters such as social learning, awareness, and ITN efficacy that help categorize SSA countries and inform country-specific strategic predictions behind the failure of achieving their GTS 2020 milestones. We focused on human behavioural interactions related to ITNs usage starting in 2000. Although ITNs were introduced in the mid-1980s decade, coverage was very low, ranging from 0 to 5\% \cite{lengeler2004insecticide}. Large-scale distribution of Long-lasting insecticidal nets (LLINs) or ITNS began only in 2005 \cite{world2013global, bertozzi2021maps}. This investigation of exploration and prediction about 38 SSA countries consists of three phases: \textbf{Phase-0 (2000-2022)}- involves parameter estimation using the Maximum Likelihood Estimation (MLE) method, while \textbf{Phase-I (2022-2025)} and \textbf{Phase-II (2025-2030)} involve optimization and projection to predict potential achievement in reducing the cases indicated as GTS milestones 2025 and 2030. We optimized the underlined cost in controlling malaria elimination, such as diagnosis and treatment of malaria cases, social awareness campaigns to promote the use of bed nets, economic support to boost daily productivity, and enhancing the efficacy of bed nets in achieving the milestones \cite{andrade2022economic}. Additionally, we apply \textit{K-Means} clustering to analyze the successes and failures of GTS milestones from various ecological-epidemiological perspectives, enriching our understanding of the trajectory towards GTS goals for all 38 SSA countries.

\section*{Results}

\subsection*{Data Analysis} 
\begin{figure}[H]
     \centering
     \includegraphics[width =1\textwidth]{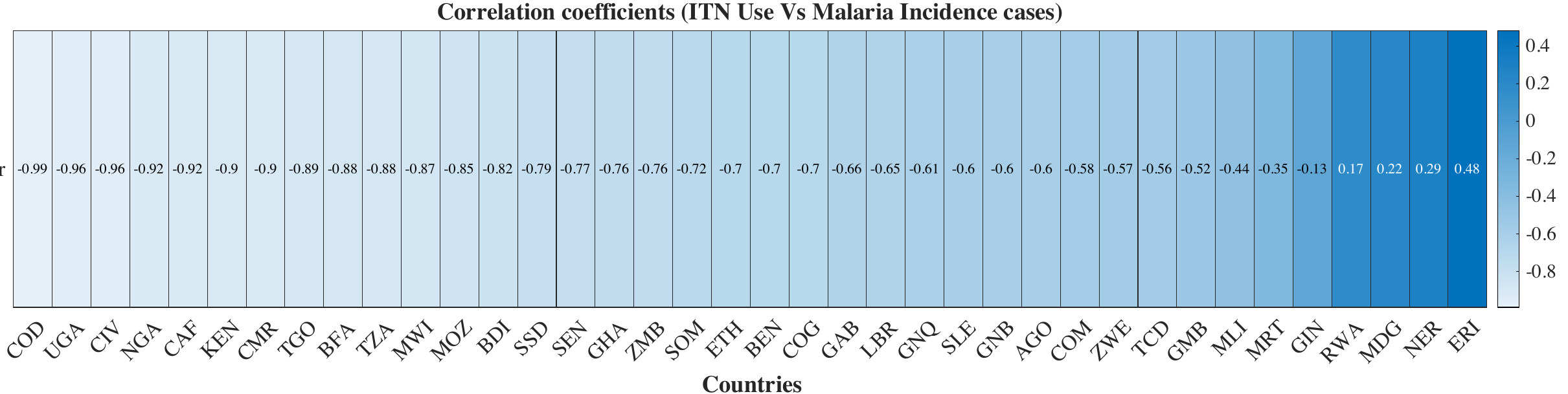}
     \caption{Heat map representing the Correlation Coefficient  between ITN use and malaria incidence cases for all 38 countries.}
\label{fig:corr_coeff}
 \end{figure}

To analyze the relationship between malaria cases and ITN usage in all 38 countries, we apply three statistical methods: correlation coefficient analysis, Granger causality, and transfer entropy to gauge causality. In the 38 countries with the highest burden, higher levels of ITN usage were generally associated with fewer malaria cases during the study period, as shown in Figure \ref{fig:corr_coeff}. This figure illustrates that the Pearson correlation coefficient ($r$) between ITN usage and malaria incidence is close to -1 for most of the countries. This negative correlation suggests that proper distribution and utilization of ITNs may have a beneficial impact on malaria prevention. Additionally, the regression line (dotted) in scatter plots (as shown in figures S2 and S3 in Supplementary Information) with negative slopes for the 38 countries supports this observation, consistently demonstrating the same negative correlation between ITN use and malaria incidence.

\begin{figure}[H]
     \centering
     \includegraphics[width =1\textwidth]{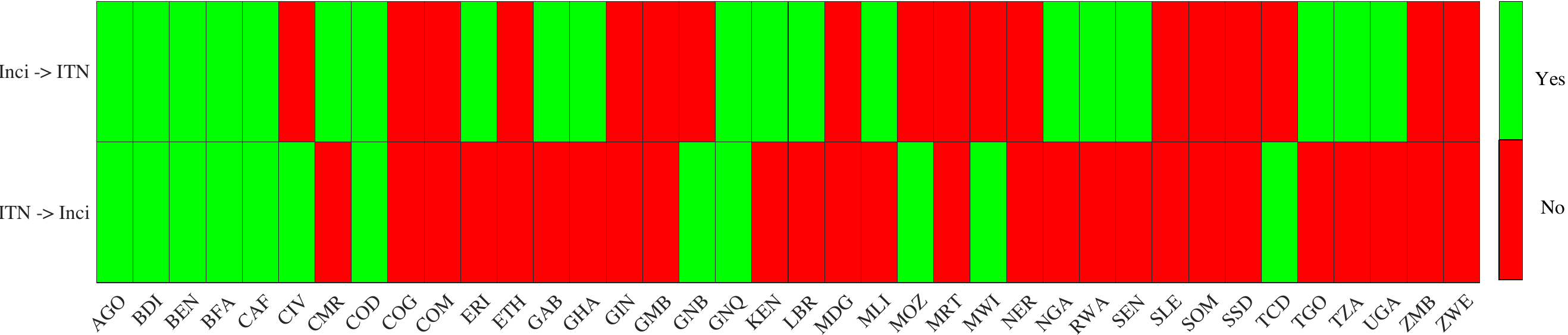}
     \caption{Heat map representing the Granger causality relationships of 38 countries. The green colour indicates potential causal relationships, while the red colour denotes the absence of causality.}
\label{fig:Granger_combine}
 \end{figure}

Delving deeper into the dynamics, the Granger causality analysis revealed a mutual feedback loop. Changes in ITN usage appeared to cause Granger-cause changes in malaria incidence, emphasizing the positive impact of ITNs. In contrast, changes in malaria incidence also appeared to have Granger-cause changes in ITN usage, indicating that an increase in malaria cases can stimulate greater ITN utilization (Figure \ref{fig:Granger_combine}). We found that 13 countries exhibited evidence that malaria prevalence Granger-causes ITN usage, but not the other way around. Five countries showed that ITN usage Granger-causes malaria prevalence, but not vice versa. Thirteen countries did not provide evidence of Granger causality in either direction, while seven countries demonstrated bidirectional Granger causality, where malaria prevalence and ITN use Granger-cause each other. Altogether, 67\% of countries exhibit that either malaria incidence (Granger) causes ITN usage or vice versa. This causative relationship strengthens the argument that ITN usage plays an important role in reducing malaria cases. The results from the Granger causality test thus corroborate the findings from the Pearson correlation analysis, providing a more comprehensive understanding of the dynamic relationship between ITN usage and malaria incidence and vice versa.

However, both correlation and Granger causality are based on linear models; therefore, to capture the anticipated nonlinear interdependency of the data series, we estimated the transfer entropy using the Binning method according to the uniform estimation approach between malaria incidence cases and ITN use (a detailed description of the methods used to estimate the probabilities involved in the evaluation of the TE is given in the Supplementary Information S2.3).  Figure \ref{fig:TE} shows a discernible higher entropy flow of incidence $\rightarrow$ ITN compared to ITN$\rightarrow$Incidence. This suggests that a greater amount of information flows from the past (and present) states of malaria incidence to predict future values of ITN usage.
 \begin{figure}[H]
     \hspace{0.5cm}
     \includegraphics[width =1.5\textwidth]{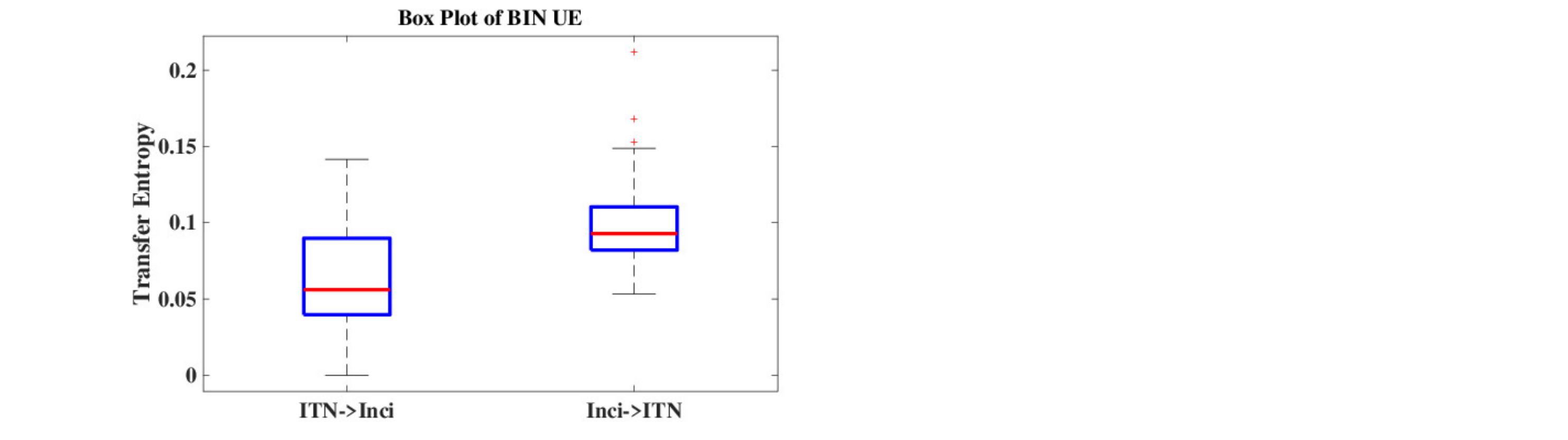}
     \caption{Box plot illustrating the summary of transfer entropy estimated using the Binning method with Uniform Estimation (BIN UE) between two variables: ITN usage and malaria incidence (Inci), across 38 countries in Sub-Saharan Africa. A higher value of transfer entropy implies that more information flows from malaria incidence to predict future values of ITN usage, compared to the information flowing from ITN usage to predict future values of malaria incidence (a detailed description is given in Supplementary Information).}
\label{fig:TE}
 \end{figure}
These statistical results underscore the positive impact of ITNs on reducing malaria transmission within communities, a critical but challenging outcome for many Sub-Saharan African countries. The effectiveness of ITNs depends not only on their distribution but also on individual behavioural factors influencing their use, both of which are crucial for advancing malaria control efforts \cite{williams2004critical, monroe2021improving}. Motivated by these findings, we investigated deeper into the issue by developing a mechanistic model to analyze and project key factors contributing to the shortcomings. Our goal is to identify actionable strategies for addressing these challenges and achieving the future milestones outlined in the GTS. \\ 
In the next section, we fit the integrated behaviour-incidence model (detailed description of model S8 is provided in Supplementary Information) to the time series of malaria incidence cases (per 1,000) and ITN use (in $\%$) from 2000-2022 for all 38 countries of SSA to project the pattern of malaria incidence cases and ITN usage from 2022 to 2025 and further to 2030 for all 32 countries except for the six countries—Niger, Benin, Mali, Rwanda, Madagaskar, and Eritrea—that showed a lack of good fit (see Supplementary Information for detailed discussion). 
\subsection*{Phase-0: Estimation and Sensitivity Analysis}
\begin{figure}
     \centering
 \includegraphics[width=1\textwidth]{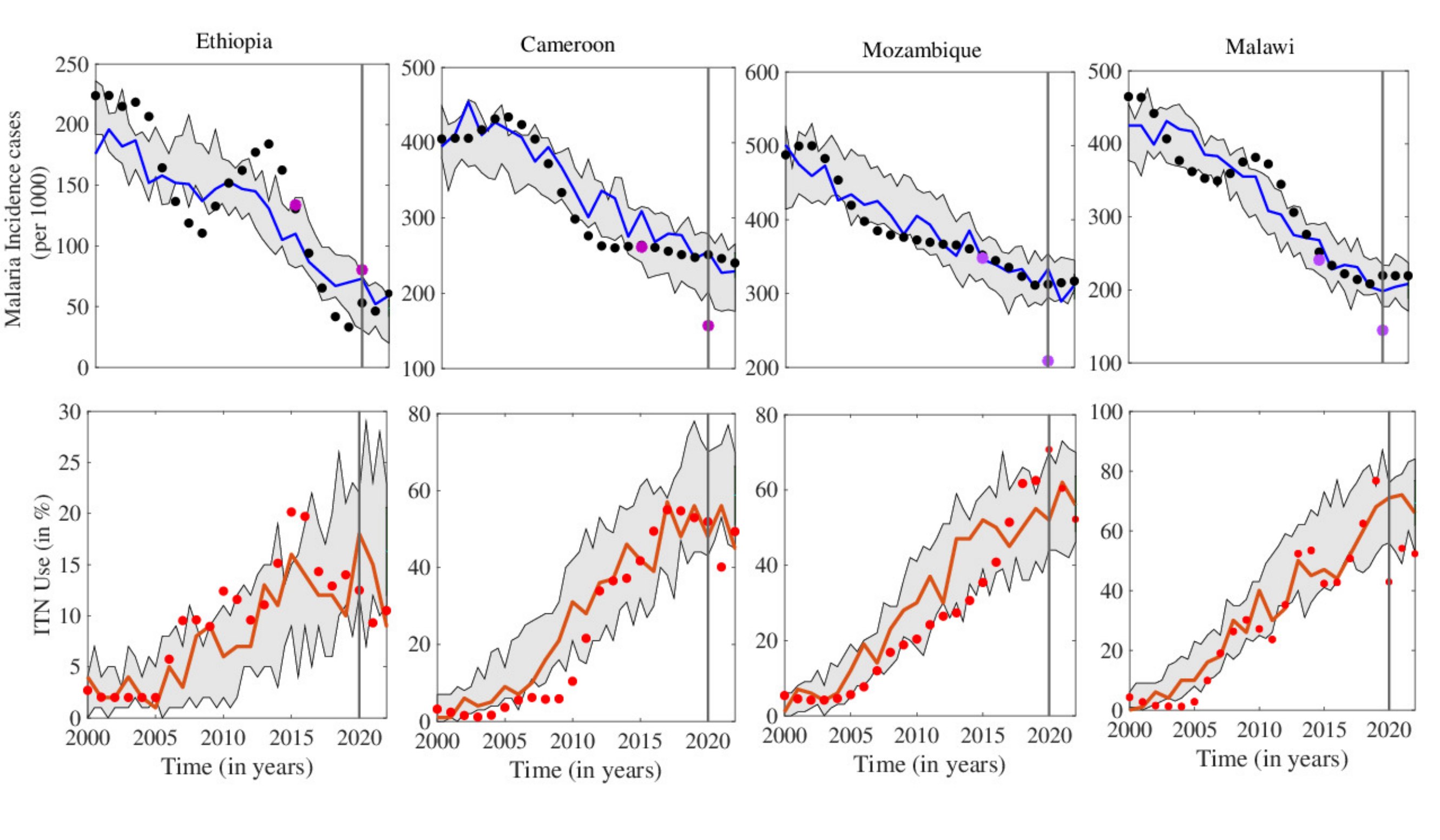}
     \caption[The behaviour-incidence model fitting to the reported malaria cases and ITN use using MLE]{The behaviour-incidence model fits to the reported cases and ITN usages for four countries (a) Ethiopia, (b) Cameroon, (c) Mozambique and (d) Malawi. The black and red dots represent reported malaria cases and ITN usage data, respectively. The blue and red lines correspond to model predictions. The grey part indicates the confidence interval determined using the bootstrap method (details are provided in sections S3 and S6 in the Supplementary Information).}
     \label{fig:Phase-0}
 \end{figure}

\begin{figure}
    \hspace{-1cm}
    \includegraphics[width=1.2\linewidth]{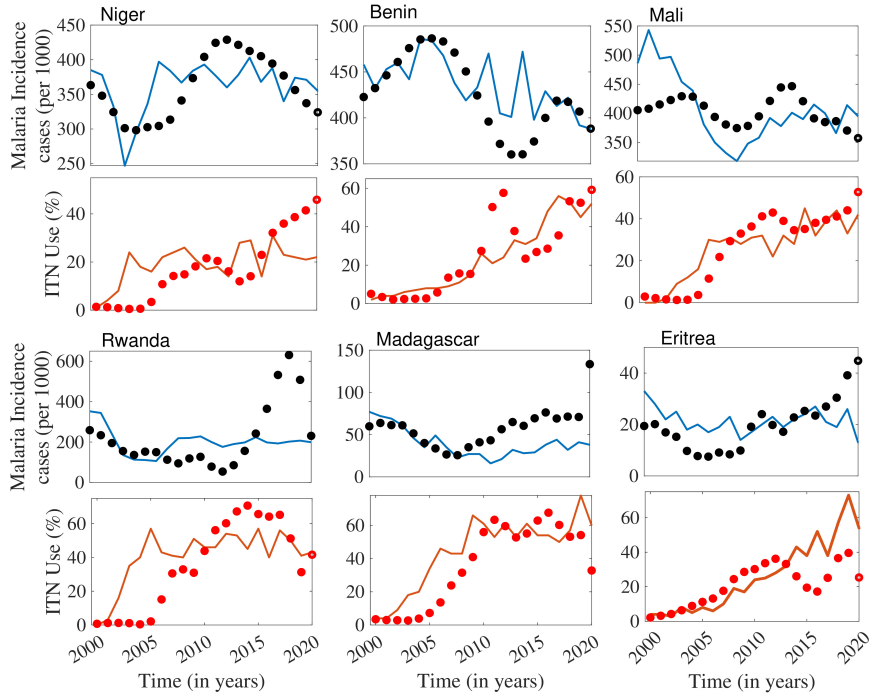}
    \caption[Dynamics of our behaviour-incidence model fitted predictions and empirical patterns of incidence and ITN use from 2000-2020 of six countries]{Dynamics of our behaviour-incidence model fitted predictions and empirical patterns of incidence and ITN use from 2000-2020 of six countries. Year `0' stands as the year `2000'. Black dots represent the reported malaria cases, red dots are the reported ITN use, and blue and red curves are the model output for cases and ITNs usage, respectively.}
    \label{fig:badfit}
\end{figure}

We begin parameter estimation by fitting the integrated behaviour-incidence model for malaria incidence cases (per 1,000) and ITN use (in percentage) from 2000-2022 for all 38 countries of Sub-Saharan Africa. A detailed discussion of the estimation methodology is provided in Section S3 of the Supplementary Information. We estimate three key parameters: ($\beta, \kappa, r$), which are related to infection transmission and human behaviour regarding ITN usage. The integrated behaviour-incidence model demonstrates a very good fit to both data sets in 30 countries, with a goodness-of-fit exceeding 0.5, capturing essential non-linear dynamics reflected in malaria incidence and ITN usage during the period 2000-2022 (Figure \ref{fig:Phase-0}). A moderate fit is observed for two countries, South Sudan and Guinea, with goodness-of-fit values between 0.4 and 0.5. However, the behaviour-incidence model showed a slight lack of good fit for the remaining six countries, where the goodness of fit was below 0.4 (see Figure \ref{fig:badfit}). We excluded those six countries from further analysis throughout this paper. The estimated parameters and the goodness of fit values for the 32 countries are listed in Table S3 given in Supplementary Information.
 
 \subsection*{Phase-I \texorpdfstring{$\&$}{and} Phase-II: Optimization \& Projection }
\begin{figure}[]
    \vspace{-0.2cm}
    \includegraphics[width=1\textwidth]{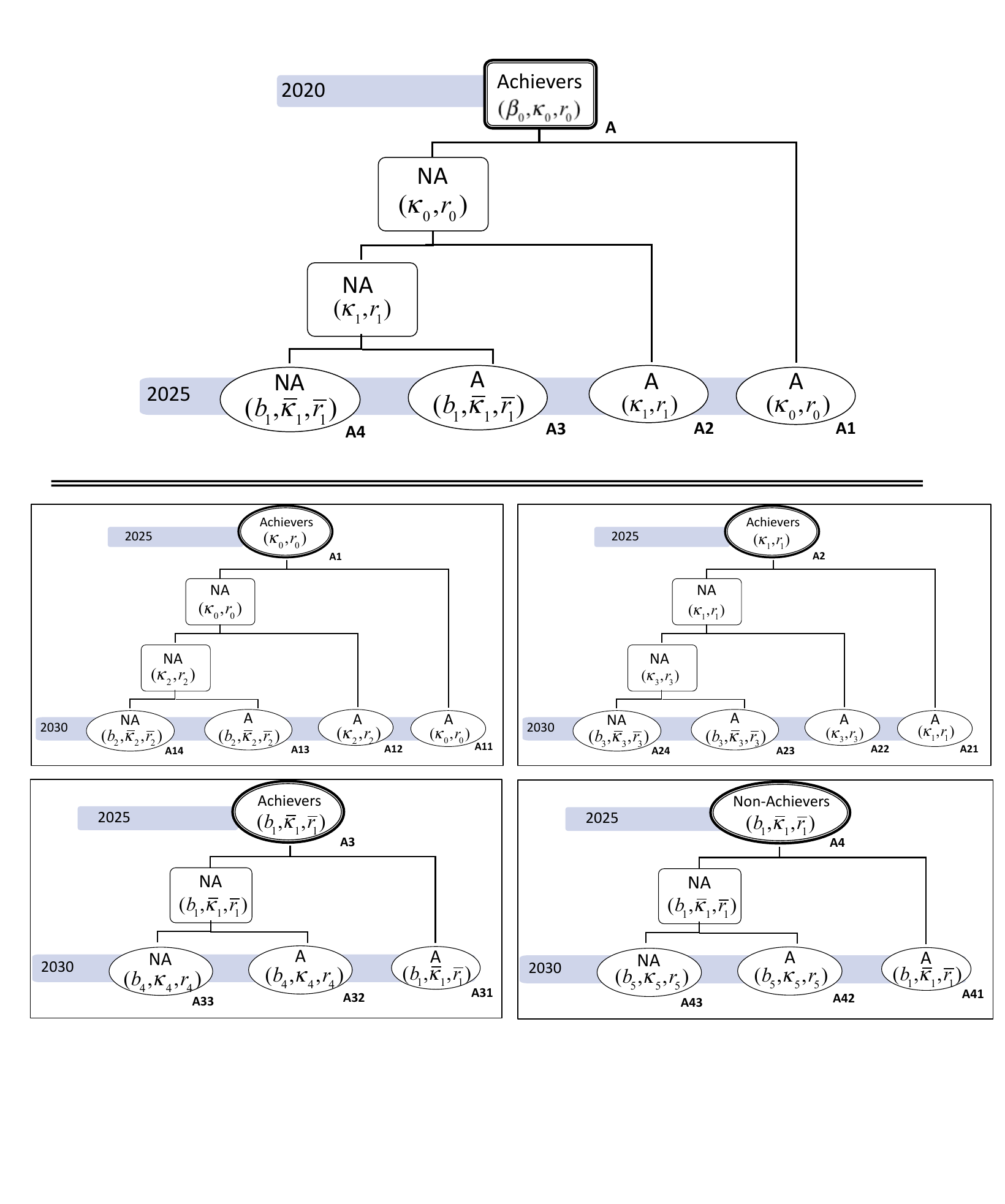}
    \vspace{-2.5cm}
    \caption[Strategic planning tree for 2020 GTS milestones achievers countries for 2025 \& 2030 milestones]{Strategic planning tree for 2020 GTS milestone achiever countries for 2025 \& 2030 milestones. \textbf{A} stands for Achiever, and \textbf{NA} stands for non-achiever. The upper panel above the horizontal line shows the planning for 2025 milestones, and the lower panel describes the planning for 2030, based on outcome at the end of 2025, either of \textbf{A1, A2, A3} or \textbf{A4}. For detailed description of this figure, see the main text.}
    \label{fig:GTSpathAchiev}
\end{figure}

\begin{figure}[]
    \vspace{-0.5cm}
    \includegraphics[width=1\textwidth]{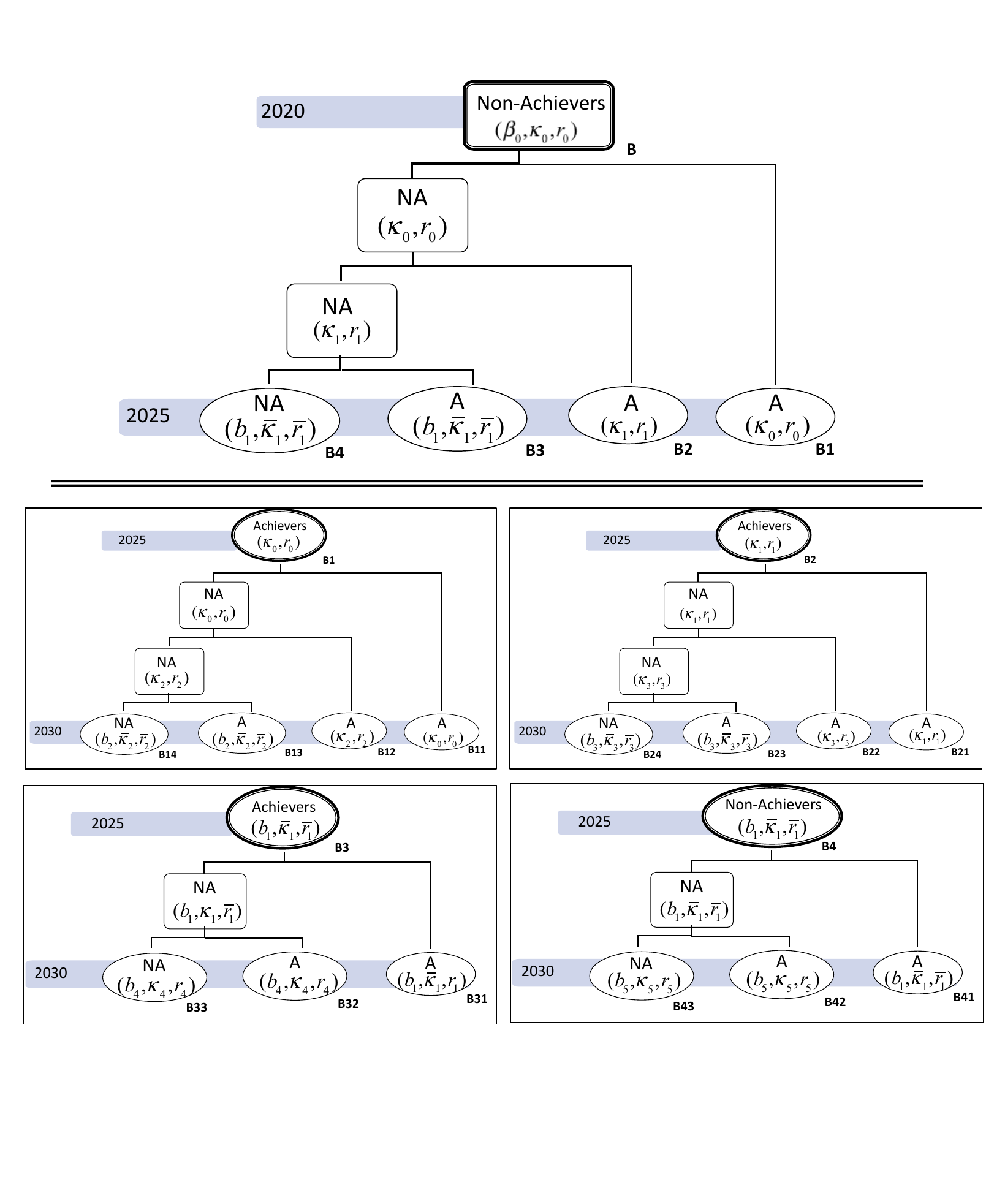}
    \vspace{-2.5cm}
    \caption[Strategic planning tree for 2020 GTS milestones Non-achievers countries for 2025 \& 2030 milestones]{Strategic planning tree for 2020 GTS milestones Non-achievers countries for 2025 \& 2030 milestones. As described earlier, the upper panel above the horizontal line shows the planning for 2025 milestones, and the lower panel describes the planning for 2030, based on outcome at the end of 2025, either of \textbf{B1, B2, B3} or \textbf{B4}. For detailed description of this figure, see the main text.}
    \label{fig:GTSpathNonAchiev}
\end{figure}

During Phase-I and Phase-II, we identify the optimal values for the estimated parameters such as the imitation rate ($\kappa$), the relative benefit of improper use of ITNs ($r$), and the effectiveness of ITNs ($b_0$)  to propose the most effective strategies for malaria control in 32 SSA countries. Although numerous country-specific factors contribute to the success or failure of GTS milestones, our focus is on reducing malaria transmission and control through ITN use and associated human behaviour. Figures \ref{fig:GTSpathAchiev} and \ref{fig:GTSpathNonAchiev} illustrate the decision-making process for developing an optimal strategic plan for each country. It is important to note that the transmission parameter $\beta$ remains fixed at the estimated value ($\beta_0$) throughout the optimization process in both Phase-I and Phase-II. For example, in Figure \ref{fig:GTSpathAchiev}, we evaluate whether countries that achieved the 2020 milestone (Group \textbf{A}) can also achieve the 2025 milestone using the same initial parameter estimates ($\kappa_0$, $r_0$). If not, we proceed by optimizing these parameters to obtain values ($\kappa_1$, $r_1$). If further optimization is needed, we adjust the ITN efficacy ($b_0$), resulting in an optimal set of values ($b_1$, $\bar{\kappa_1}$, $\bar{r_1}$). This process leads to four possible outcomes for the 2025 milestone in Group A: (i) Achiever with ($\kappa_0$, $r_0$) (path \textbf{A1}), (ii) Achiever with ($\kappa_1$, $r_1$) (path \textbf{A2}), (iii) Achiever with ($b_1$, $\bar{\kappa_1}$, $\bar{r_1}$) (path \textbf{A3}), or (iv) Non-achiever with ($b_1$, $\bar{\kappa_1}$, $\bar{r_1}$) (path \textbf{A4}). For non-achiever countries (Group \textbf{B}), a similar analysis is conducted (Figure \ref{fig:GTSpathNonAchiev}), producing four possible outcomes for 2025: \textbf{B1}, \textbf{B2}, \textbf{B3}, \textbf{B4}. In Phase-II, we follow the same approach for the 2030 milestones. Beginning with 2025 outcomes, either \textbf{Ai}'s or \textbf{Bi}'s, further optimization generates subsequent pathways for achieving or not achieving the 2030 milestone. This result in pathways for Group \textbf{A}: \{\textbf{A11}, \textbf{A12}, \textbf{A13}, \textbf{A14}\}, \{\textbf{A21}, \textbf{A22}, \textbf{A23}, \textbf{A24}\},\{\textbf{A31}, \textbf{A32}, \textbf{A33}\},\{\textbf{A41}, \textbf{A42}, \textbf{A43}\}, and for Group \textbf{B}: \{\textbf{B11}, \textbf{B12}, \textbf{B13}, \textbf{B14}\}, \{\textbf{B21}, \textbf{B22}, \textbf{B23}, \textbf{B24}\},\{\textbf{B31}, \textbf{B32}, \textbf{B33}\},\{\textbf{B41}, \textbf{B42}, \textbf{B43}\}.\\

In the Global Technical Strategy (GTS) for malaria elimination, the World Health Organization (WHO) recommended adopting and scaling up various strategies to enhance effectiveness and reduce malaria-related deaths \cite{WHO2018}. A key focus is integrating malaria prevention (vector control), diagnosis, and treatment into universal health coverage \cite{patouillard2017global, winskill2019prioritizing}. Our analysis assumes costs in four main categories to achieve the milestones set for 2025 and 2030: (i) diagnosis and treatment of malaria cases; (ii) social awareness campaigns to promote the use of bed nets; (iii) economic support to boost daily productivity, such as providing fishing nets and fencing materials; and (iv) enhancing the efficacy of bed nets. While other vector control methods, such as chemoprevention, exist, they are not included in this analysis. We optimized the total costs across these four categories over a five-year period to meet the GTS milestones. \\

The cost function to be minimized during optimization is given by 
\begin{equation}\label{eqn:GTScost}
J(b, \kappa, r) = \int_0^5\bigg[\eta_1\lambda_hS_h+\eta_2 (\kappa-\kappa_0)^2+\eta_3 (r-r_0)^2+\eta_4 (b-b_0)^2\bigg]dt,
\end{equation}
where $\eta_2$, $\eta_3$ and $\eta_4$ are per unit cost for organizing social awareness campaigns ($\kappa$), providing economic assistance ($r$) and improving the bednet efficacy ($b$). The term $\eta_1$ represents the per unit cost for every new infection ($\lambda_h S_h$), factoring in treatment costs, lost wages, and even disease-induced deaths. We may mention here that $\kappa_0$ and $r_0$ represent the status of efforts in social awareness campaigns and economic assistance estimated at the year 2020. It is also important to note that $\eta_4$ becomes zero when there is no need to alter the efficacy of ITN ($b_0$) in order to achieve the milestone.

\vspace{0.5cm}

\begin{figure}[H]
    \hspace{-1cm}
\includegraphics[width=1.1\textwidth]{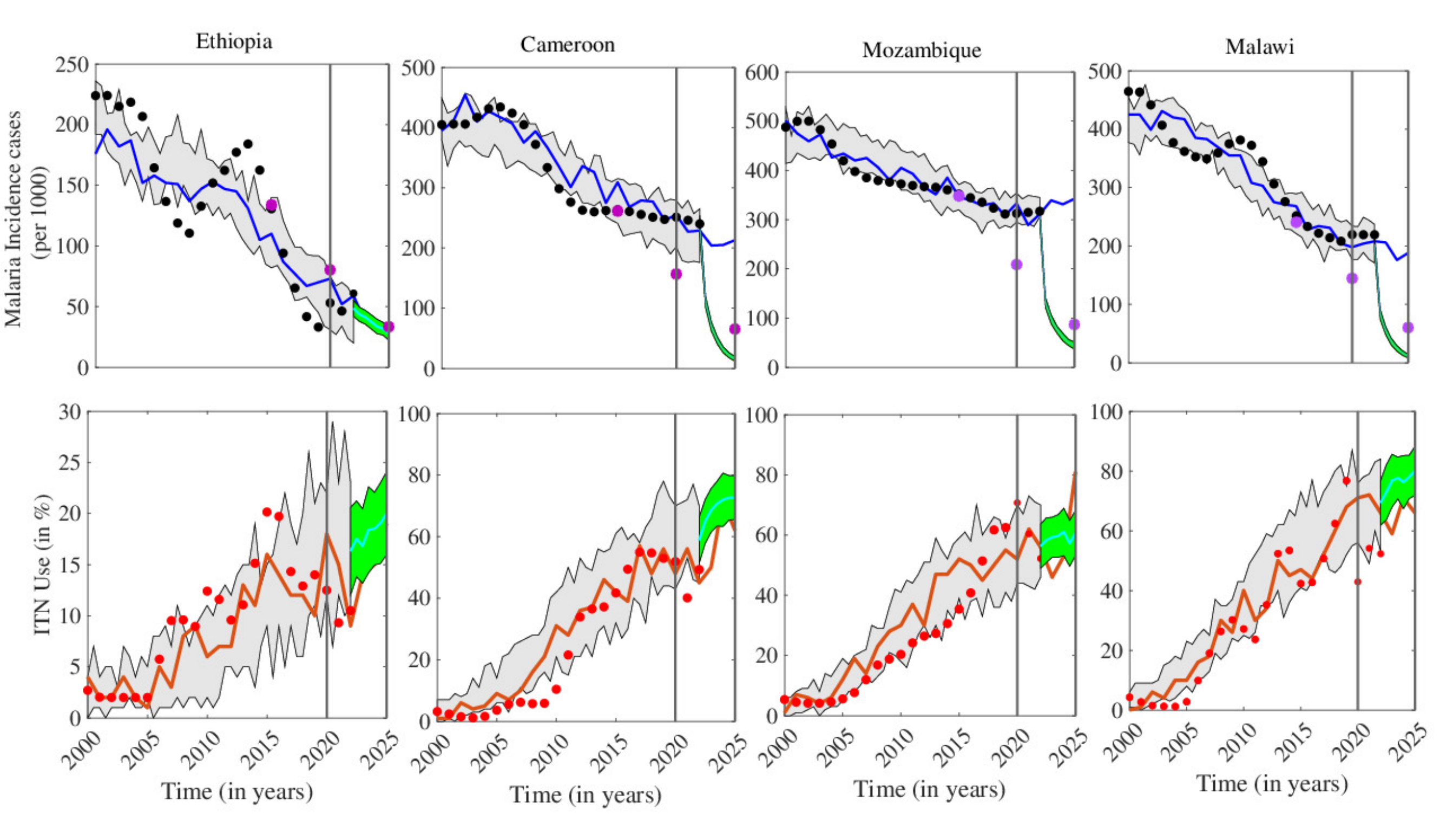}
   \caption{Optimized projections from 2020-2025 for countries (i) Ethiopia ($A_1$), (ii) Cameroon ($B_3$), (iii) Mozambique ($B_3$) and (iv) Malawi($B_3$). Labels of graphs and curves are similar to Figure \ref{fig:Phase-0}. The pink dots on the vertical line along 2020 and 2025 indicate the milestones for those specific countries. See Fig \ref{fig:ideal GTS} for more details. Path-wise projections of other countries are given in Figures S5, S6,S7 in Supplementary Information. The green part is the confidence interval determined using the bootstrap method.}
   \label{fig:Phase1}
\end{figure}

To characterize the optimal path towards achieving the 2025 milestone in Phase I, we extend the model's projection to 2025 using estimated values $(\kappa_0, r_0)$ obtained in 2020. If a country fails to meet the milestone, we employ Latin hypercube sampling (LHS) to perform the optimization, minimizing the cost function given by equation (\ref{eqn:GTScost}). Initially, we seek the optimal values $(\kappa_1, r_1)$ by setting $\eta_4 = 0$ in equation (\ref{eqn:GTScost}), to determine if the milestone can be achieved without adjusting ITN efficacy $b_0$. If the milestone still remains unattained, we consider $\eta_4 \neq 0$, and optimize for the set ($b_1$, $\bar{\kappa_1}$, $\bar{r_1}$). We find that all four countries that successfully achieved the 2020 milestone can also achieve the 2025 milestone without changing ITN efficacy ($b_0$). However, two of these achiever countries—Gambia and Ghana—require a higher imitation rate($\kappa_1$), and/or a reduced relative benefit ($r_1$) to meet the target. On the other hand, there are 2020 non-achiever countries—such as Cote D’Ivoire, Comoros, Burundi, Guinea-Bissau, Chad, Gabon, Sierra Leone, Liberia, Guinea, Mozambique, Zambia, and Angola—that need improvements in ITN efficacy ($b_0$) along with higher imitation rates($\kappa_1$) and/or lower relative benefit ($r$) to achieve the 2025 milestone (see Table S8 for the optimized value of parameters). Figure \ref{fig:Phase1} demonstrates the projections of malaria incidence cases and ITN usage in Ethiopia, Cameroon, Mozambique, and Malawi from 2020 to 2025.

\begin{figure}
    \hspace{-1cm}
    \includegraphics[width=1.1\textwidth]{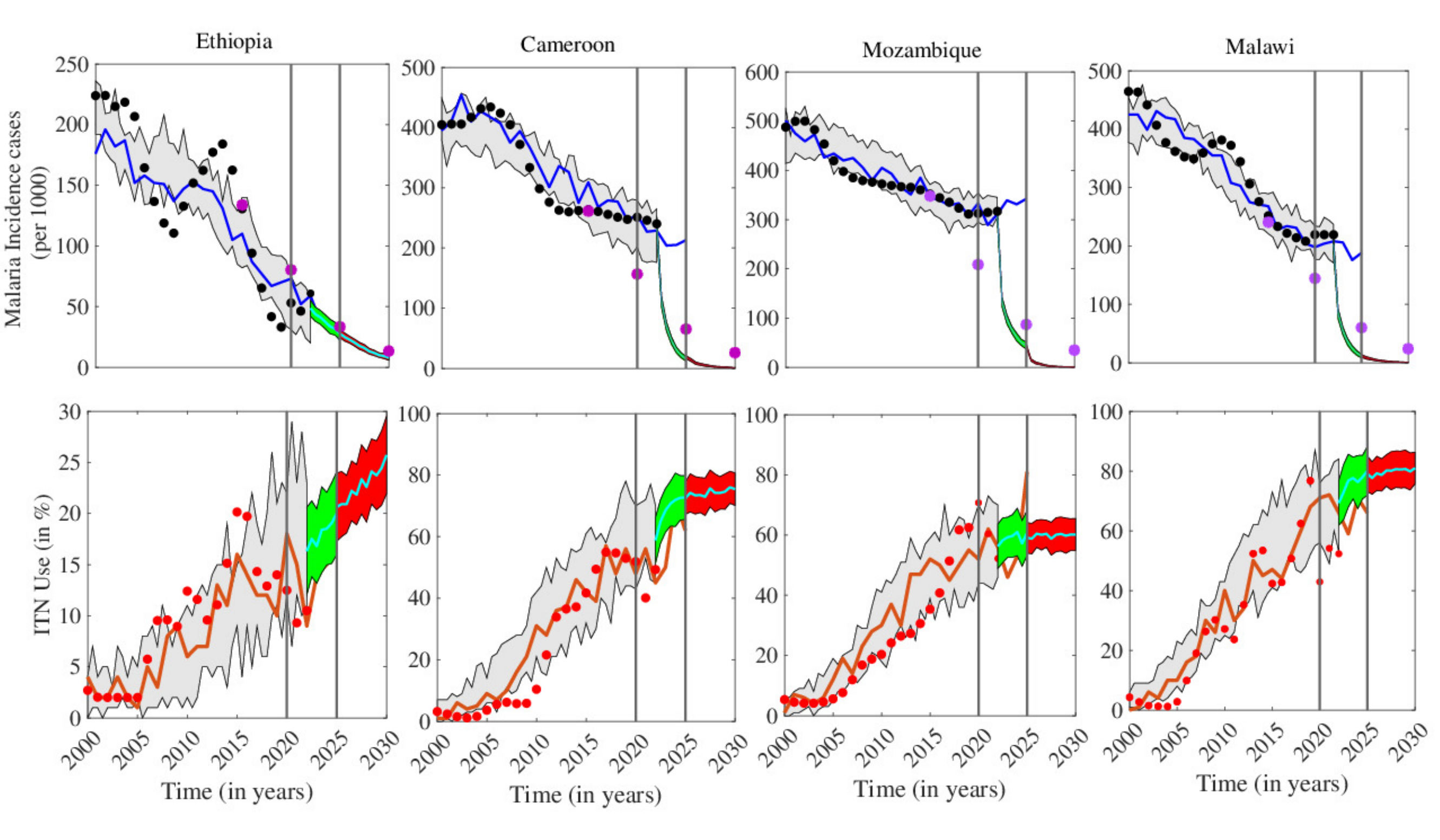}
    \caption{Optimized projections from 2025-2030 for countries (i) Ethiopia ($A_{11}$), (ii) Cameroon ($B_{31}$), (iii) Mozambique ($B_{32}$) and (iv) Malawi ($B_{31}$). The green part in 2025-30 is the confidence interval determined using the bootstrap method. Path-wise projections of other countries  are given in Figures S5, S6, S7 in Supplementary Information}
    \label{fig:phase2}
\end{figure}

In Phase-II, the behaviour-incidence model's projection is extended to 2030, using the estimated optimized values of ITN efficacy, imitation rates $\kappa$, and relative benefit ($r$) from 2025. Our findings indicate that countries classified as 2020-Achievers can reach the 2030 milestone without requiring further changes in the social learning rate ($\kappa$) or the relative benefit of ITN misuse ($r$) beyond 2025. This also applies to several 2020-Non-Achievers countries, except for the Democratic Republic of Congo, Togo, Malawi, Cameroon, and Uganda. These countries, which did not require changes in ITN efficacy ($b_0$) in the 2025 projections, will need improvements in ITN efficacy to meet the 2030 goal. In addition, some countries that needed adjustments in the efficacy of ITN in the 2025 projections—Côte d'Ivoire, Comoros, Burundi, Guinea-Bissau, Chad, Gabon, Sierra Leone, Liberia, and Guinea—are projected to achieve the target of 2030 without further changes to their strategies. However, three of the 2020 non-achieving countries—Mozambique, Zambia, and Angola—will require additional improvements in social learning rate ($\kappa$), relative benefit ($r$) and ITN efficacy ($b_0$) to reach the milestone of 2030.
Figure \ref{fig:phase2} demonstrates the optimal projections of malaria incidence and ITN use, while Table
\ref{tab:strategic Paths} provides detailed, country-specific strategic plans to achieve the milestones of 2025 and 2030. 

\begin{table}[H]
\caption{Summary of strategic planning paths of all 32 countries for 2025 and 2030 milestones. $A_{ij}$ ($B_{ij}$) denotes strategic path, where $A_i$ ($B_{i}$) is required for achieving milestone in 2025, and $A_{.j}$ ($B_{.j}$) path is required for achieving milestone in 2030 (see Figures \ref{fig:GTSpathAchiev} \& \ref{fig:GTSpathNonAchiev} for references to path definitions). Countrywise targeted intervention are required to achieve milestones in 2025 and 2030, and relative increase in ITNs usage can be found in Table S4 \& S5 in the Supplementary Information}
\vspace{-2mm}
\begin{center}
\resizebox{\textwidth}{!}{
\begin{tabular}{ p{0.15\linewidth}p{0.85\linewidth}}
\hline
 \textbf{Path} & \textbf{2020 milestone achiever countries}  \\
 \hline  
 \hline
$A{11}$& Ethiopia, Mauritania \\

$A{21}$ & Gambia, Ghana\\
\hline
&\textbf{2020 milestone non-achiever countries}\\
\hline
\hline

$B{21}$ &  Burkina Faso, Central African Republic, Congo, Equatorial Guinea, Kenya, Nigeria,  Senegal,  Somalia, South Sudan, Tanzania, Zimbabwe     
\\

$B{23}$ & Togo \\

$B{31}$ & Burundi, Cameroon, Comoros, Cote D'Ivoire, Democratic Republic of Congo, Gabon,   Guinea- Bissau, Liberia,  Malawi, Sierra Leone, Uganda, Zambia\\

$B{32}$ & Angola, Chad, ,  Guinea, Mozambique\\

\hline
\end{tabular}}
\label{tab:strategic Paths}
\end{center}
\end{table}

These analyses provide valuable insights for Global Technical Strategy (GTS) planning related to malaria control and elimination strategies in SSA countries. It emphasizes that even countries that missed the 2020 milestone have the potential to make substantial progress in their malaria control programs. Public health officials can play a pivotal role  by increasing public awareness and educating communities on the proper use of ITNs through targeted campaigns and the use of social media. In addition,  governments need to introduce or expand regular subsidies aimed at improving socioeconomic conditions, such as providing alternative nets for fishing and fencing to reduce the misuse of ITNs. In some cases, enhancing the effectiveness of existing ITNs is also critical factor. Specific countries may need more frequent ITN replacements to maintain their protective efficacy, thereby improving their chances of achieving the 2025 and 2030 GTS milestones. These proactive measures could significantly strengthen the impact of malaria control initiatives, aligning SSA countries with global targets.

\begin{figure}
    \hspace{-0.1cm}
    \includegraphics[width=1\linewidth]{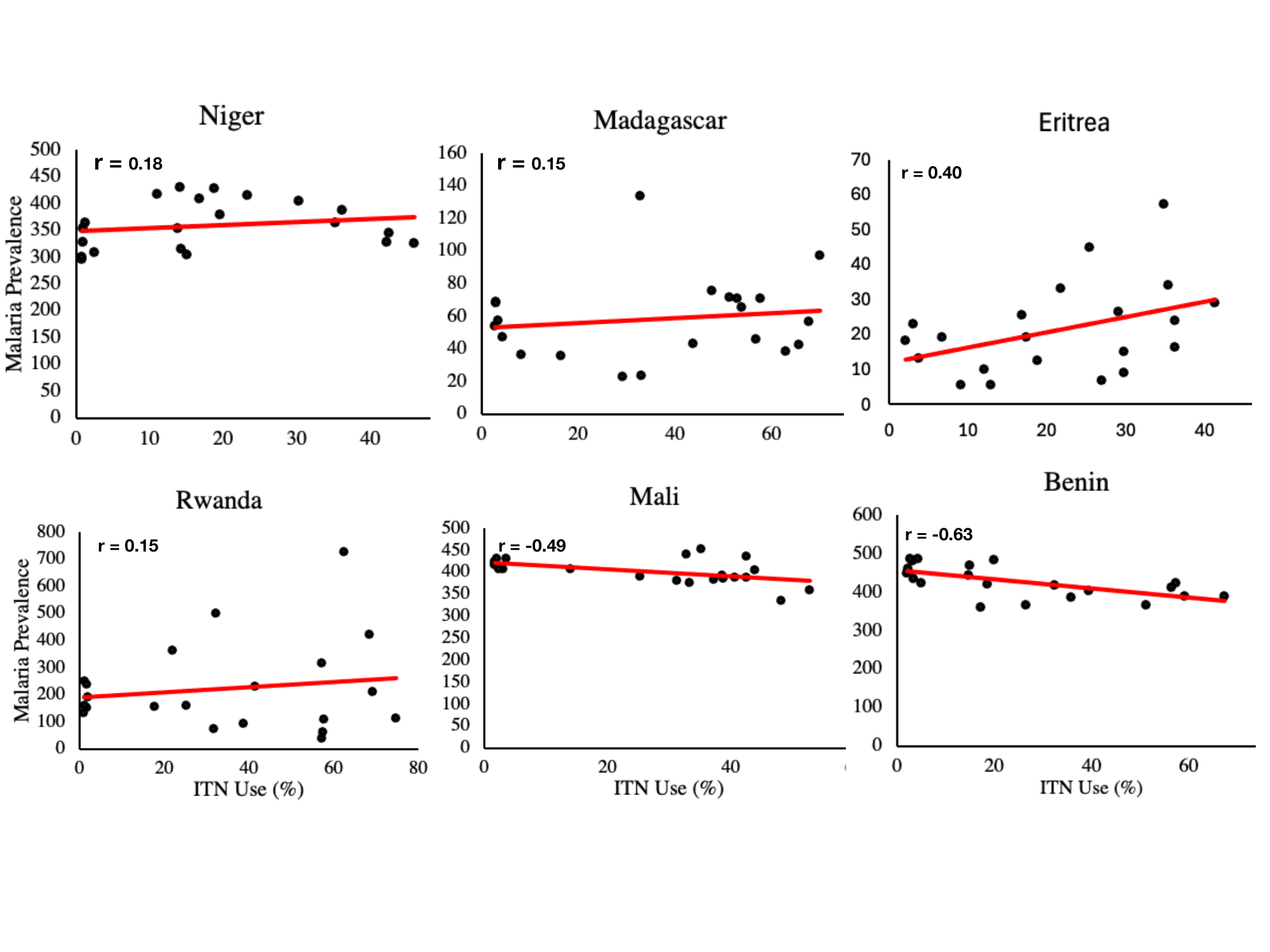}
    \vspace{-1cm}
    \caption[Correlation between empirical malaria prevalence and ITN usage in outliers Niger, Rwanda, Mali, Madagascar, Eritrea, and Benin.]{Correlation between empirical malaria prevalence and ITN usage in outliers Niger, Rwanda, Mali, Madagascar, Eritrea, and Benin.}
    \label{fig:corr_badfit}
\end{figure}

\subsection*{K-means Clustering}
We used the calibrated and estimated parameter values in the pre-estimation and estimation exercises and performed the cluster analysis of 32 countries in Sub-Saharan Africa. These parameters are $\Lambda_h, \Lambda_v, \delta_h, \beta, \kappa, r$. The parameters $\Lambda_h$ and $r$ here represent the demographic and socioeconomic characteristics of the nation, while $\Lambda_v$ and $\beta$ represent the growth of the vector population and the per capita mosquito biting rate, respectively, factors that indicate the transmission potential. The mortality rate $\delta_h$ accounts for the effectiveness of malaria control measures and public health interventions in each country. Using the scree plot in Figure S4(a) in Supplementary Information, we identified three components -$(\Lambda_v, \beta, r)$ that have eigenvalues greater than $1$, and further based on factor analysis, we determine that three principal components -$(\Lambda_v, \beta, r)$ -play a important role in explaining the variance across countries regarding their malaria elimination status. The Elbow method further indicates that the optimal number of  clusters is five, with a varying number of countries within each cluster, as shown in Figure S4(b) in Supplementary Information. Figure \ref{fig:k-cluster} shows the cluster of countries, and tables S6 and S7 provide the list of countries within five primary clusters, together with the corresponding mean values $(\Lambda_v, \beta, r)$ for each cluster, respectively. This cluster analysis offers a nuanced understanding of the factors that influence malaria control and can help tailor interventions to the specific characteristics of each group of countries. 

\begin{figure}
    \hspace{-1cm}
    \includegraphics[scale=0.45]{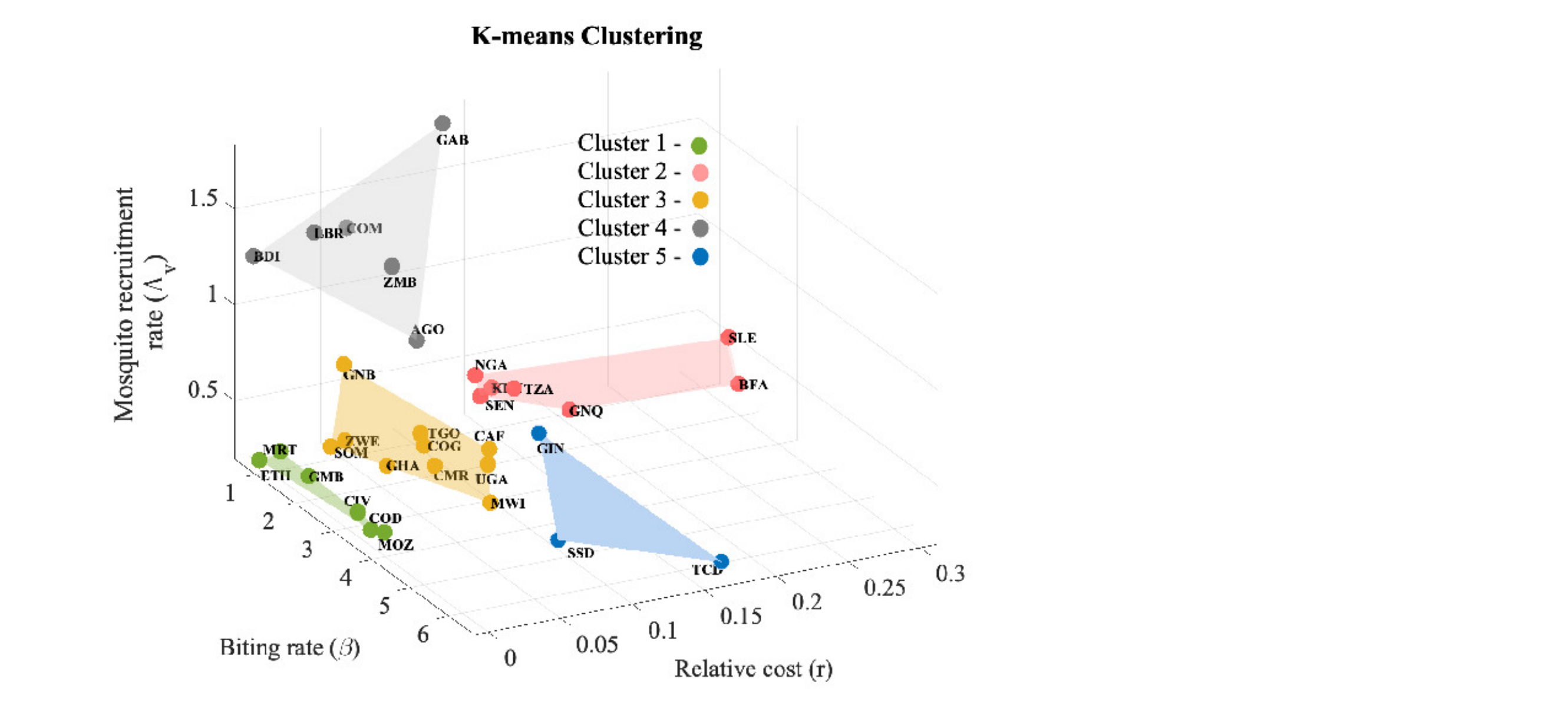}
    \caption{Clusters of all 32 countries based on parameter values obtained in the pre-estimation and estimation exercises. The figure indicates the difference in and variance within the clusters in three components: mosquito biting rate ($\beta$), relative cost ($r$), and mosquito recruitment rate ($\Lambda_v$). For detailed analysis, see the main text.}
    \label{fig:k-cluster}
\end{figure}

\begin{figure}
    \hspace{-0.1cm}
    \includegraphics[trim = 0cm 0cm 0cm 0.2cm , clip, width=1\linewidth]{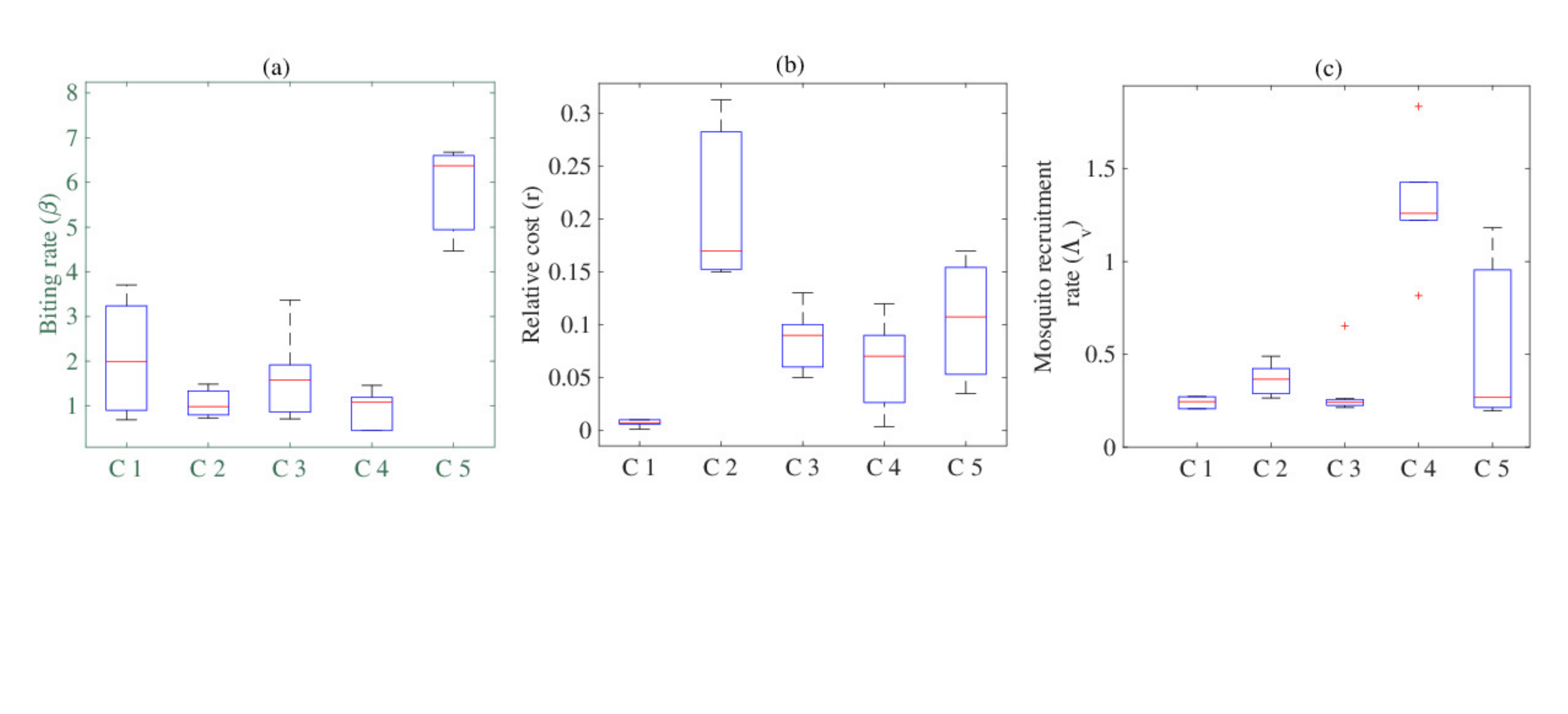}
    \caption{Distribution of (a) biting rate, (b) relative cost in improper use of ITNs, and (c) per capita mosquito recruitment rate across clusters.}
    \label{fig:boxplot}
\end{figure}

The variability in the principal components across clusters is illustrated in Figure \ref{fig:boxplot}. The per capita mosquito bite rate ($\beta$) is notably highest in Cluster 5, while it is lowest in Clusters C2, C3, and C4, with a moderate level observed in Cluster 1 (Figure \ref{fig:boxplot}(a)). Conversely, the per capita mosquito recruitment rate is highest in Cluster 4, exhibits discernible variability in Cluster 5, and shows moderate levels with minimal variation in Clusters 1 to 3 (Figure \ref{fig:boxplot}(b)). These findings imply that the likelihood of potential infection transmission is higher in countries belonging to Clusters 1 and 5. In particular, countries such as Chad, Guinea, and South Sudan, which are classified within these clusters, exhibit relatively higher transmission rates between SSA countries. Furthermore, countries in Cluster 1 may also face moderate variability in transmission rates, indicating that targeted malaria control measures may be necessary in these regions to address the elevated risk of transmission.\\
Another key component, the relative benefit of improper use of ITNs ($r$), serves as an indicator of the socioeconomic conditions in the countries and their influence on ITN (mis)use. As depicted in Figure \ref{fig:boxplot}(c), the relative benefit of improper ITN usage is moderate to low across all clusters, except for Cluster 4. Countries in Cluster 5—including Chad, South Sudan, and Guinea—either face poorer economic conditions and/or have lower awareness and knowledge about ITN usage. This scenario is associated with higher levels of ITN misuse in these nations. Overall, this analysis provides valuable insights for policymakers, enabling them to tailor effective malaria control and elimination strategies in these specific regions.
 
\section*{Lack of Good Fit to the Behaviour-Incidence Model in Six Sub-Saharan Countries}

The behaviour-incidence model failed to capture malaria incidence and ITN usage patterns in six Sub-Saharan African countries—Niger, Rwanda, Mali, Benin, Madagascar, and Eritrea—out of 38 analyzed. These discrepancies, with goodness-of-fit below 20\% (except Benin \& Mali), highlight the model’s inability to account for complex socio-economic, climatic, and healthcare-related factors.

In \textit{Niger}, malaria incidence displayed a cyclic trend, while ITN usage fluctuated inconsistently, partly due to unequal healthcare distribution and socio-political challenges. \textit{Benin} also showed poor alignment, with high ITN use undermined by low public awareness and unsuitable living conditions. \textit{Mali}’s steady malaria incidence contrasted with increasing ITN usage, largely due to diverse eco-climatic zones, decaying net efficacy, and infrastructure issues. \textit{Rwanda} initially achieved effective malaria control but faced an eight-fold surge in cases between 2012-2016, driven by climatic changes, insecticide resistance, and short net durability. \textit{Madagascar} and \textit{Eritrea} had low malaria incidences but rising trends despite consistent ITN usage, influenced by poor healthcare infrastructure and low economic productivity. For example, Madagascar’s informal economy limits progress in health system improvements. \textit{Nigeria} also exhibited issues, such as low ITN usage due to cultural beliefs, lack of awareness, and logistical challenges, demonstrating that economic strength alone cannot ensure effective ITN utilization. We have discussed in detail country-specific issues in the Supplementary Information.

These findings, however, underscore the limitations of the simplified assumptions in the behaviour-incidence model and the need for  incorporating more nuanced, context-specific variables to improve its accuracy and applicability.

\section*{Discussion and Limitations}
Within the strategic framework of Global Technical Strategies (GTS) for malaria, the WHO African Region emphasizes the need for Sub-Saharan African (SSA) countries to maximize the impact of existing lifesaving interventions. This effort aims to reduce the incidence and mortality of malaria while accelerating progress toward elimination. The primary objective of national malaria programs in areas with moderate to high transmission in SSA is to ensure equitable access to insecticide-treated nets (ITNs) and implement other strategies to improve vector control \cite{WHO2018}.  

This analysis aimed to comprehensively evaluate the successes and shortcomings of GTS implementation in 32 SSA countries, focusing on ITN usage and vector control efforts. Our findings suggest that several countries with poor performance could substantially improve their outcomes with additional targeted initiatives. Our analysis focuses on three targeted initiatives: improving ITN campaigns to raise social awareness about their usage, providing economic assistance, and improving the efficacy of bed nets. Although these initiatives are pivotal, it is important to acknowledge that numerous other factors can influence the success of malaria elimination efforts and the achievement of GTS goals. 

The mechanistic behaviour-incidence model presented here was instrumental in accurately capturing temporal trends in ITN usage and malaria cases in various SSA countries. Our optimized projections for 32 well-estimated countries provide a roadmap for targeted initiatives necessary to achieve the Global Technical Strategy (GTS) goals for 2025 and 2030. Although Ethiopia and Mauritania, both 2020 GTS achievers, are on track to meet the 2025 and 2030 goals with their current efforts, our findings underscore the urgent need for Gambia and Ghana to scale up their social awareness campaigns drastically. Specifically, Gambia should increase its campaign efforts nearly ninefold to stabilize ITN usage compared to its current situation. At the same time, Ghana requires a similar expansion to achieve an estimated $31.69\%$ increase in ITN usage by 2025 and 2030, effectively reducing the burden of malaria.  

These findings align with recent studies on Ghana and the Gambia \cite{doe2024ownership, klu2022mixed, tweneboah2022role, akuffo2021insecticide, jawo2022knowledge, yeboah2023caregivers}. For example, Doe et al. (2024) \cite{doe2024ownership} identified potential regional disparities in ITN ownership and usage in Ghana, highlighting protective factors such as marital status, higher educational attainment, higher income levels, and age (25 years or older). They recommended community-level campaigns promoting ITN use in conjunction with education initiatives to increase awareness. Similarly, Jawo et al. (2022) \cite{jawo2022knowledge} emphasized the importance of health education programs on malaria prevention in the Gambia, suggesting that these programs be distributed through television, radio, and other media platforms and advocating for health education as a mandatory topic in educational institutions. These insights underscore the critical role of targeted interventions and educational strategies in addressing ITN use disparities and achieving malaria control goals.

Several East and Central African countries that fell short of the 2020 milestone demonstrate the potential to meet the 2025 targets, provided that there is a remarkable increase in social awareness about the ITN campaign. To achieve this, most of these countries must dramatically scale up their social awareness efforts by five to 10 times the current levels to increase ITN usage. Currently, ITN usage in countries such as Kenya, Equatorial Guinea, Tanzania, the Central African Republic (CAR), Nigeria, and South Sudan is approximately 45\% or lower. Meeting the 2025 milestone will require a relative increase in ITN usage of more than 100\% (see Table S4). However, most of these countries would need minimal additional efforts to achieve the 2030 milestone, except Nigeria and South Sudan. The same effort to achieve the 2025 milestone should continue to achieve the 2030 milestone. This underscores the critical role of social awareness in improving ITN usage and advancing malaria control and elimination efforts.\\
Recent survey studies \cite{lacey2023combating, githinji2021preventing, margaret2022social, onyinyechi2023effectiveness} conducted in Kenya highlight that public education on prevention strategies and their adoption remains a critical challenge in the fight against malaria. These studies recommend leveraging innovative approaches such as digital tools for reporting or notification of malaria cases, maternal malaria clinical education, and community awareness programs implemented through schools and NGOs to improve malaria prevention efforts throughout the country. Similarly, recent research \cite{liheluka2023community} in the northwestern and southern regions of Tanzania, as well as other areas \cite{croke2021impact}, identified several barriers to the use of ITNs. These include socioeconomic limitations, behavioural challenges, and widespread misconceptions and beliefs regarding the use of ITN. Some studies \cite{croke2021impact} also emphasize the importance of increasing political awareness and engagement within communities. For example, a bed net distribution campaign in Tanzania demonstrated remarkable improvements in malaria prevention outcomes when communities selected leaders who prioritized malaria control, particularly in endemic regions. Furthermore, other studies \cite{mohamed2020factors, hirai2022exploring, ayeni2020factors,  makuvara2024malaria, kwansa2024women} identified a variety of additional factors—lack of knowledge, cultural beliefs, misconceptions, and sociocultural practices—that inhibit the effective use of ITN in sub-Saharan African countries, including Tanzania, Somalia, and South Sudan. In addition to awareness in society and individual perceptions about malaria, the low usage of ITN in Nigeria may be due to conducting the survey there during a low malaria transmission season \cite{storey2018associations}. However, addressing these barriers through targeted social awareness campaigns and culturally sensitive interventions is critical to improving ITN usage and achieving malaria control goals. 

Since 2005, ITNs have been a cornerstone in reducing malaria morbidity and mortality in sub-Saharan Africa. In addition, recent declines in malaria cases and deaths can be attributed to factors such as accessibility to ITN, usage rates, bioefficacy, durability \cite{hiruy2021effect, hiruy2023durability}, etc. Several studies \cite{lindsay2021threats, accrombessi2023efficacy, sahu2020evaluation, adageba2022bio, da2023evaluation, sih2024efficacy} conducted in malaria-endemic regions, including Sub-Saharan Africa (SSA), India, and Brazil, emphasize that the reduced efficacy of ITNs is a significant barrier to their widespread use and an impediment to malaria elimination efforts. Our findings align with this observation, revealing that many SSA countries, such as Togo, Liberia, Gabon, Sierra Leone, Burundi, the Democratic Republic of Congo (DRC), Côte d'Ivoire, Uganda, Angola, and Zambia, must prioritize improving ITN effectiveness to meet malaria milestones set for 2025 or 2030. This effort should be complemented by expanding social awareness campaigns. In particular, countries such as Angola, Chad, and Uganda, where current ITN usage is below 40\%, need to adopt multifaceted strategies to align with the goals of the Global Technical Strategy for Malaria (GTS) (see Figure S5). Strengthening ITN performance while addressing barriers to adoption will be critical to advancing malaria elimination in these regions.\\

Our analysis has certain limitations. Although we used yearly data on malaria incidence and ITN use between 2000 and 2020, higher-resolution data could improve predictive accuracy. The aggregates for years may not fully capture seasonal or monthly variations in malaria transmission, which are affected by factors such as rainfall, temperature, and vector breeding cycles. Furthermore, annual data may overlook short-term behavioural changes or interventions that could potentially influence ITN usage and malaria incidence over shorter time frames. This temporal limitation may obscure rapid changes and hinder the evaluation of the effectiveness of the intervention during brief periods. Furthermore, in countries where surveys are conducted regularly, data collection should ideally occur during the same season each year. Given the seasonal variations in ITN usage, season-specific estimates could improve predictions and aid in developing more effective country-specific actionable strategies.\\
The assumption that the efficacy of ITN remains constant throughout the study period limits the predictive potential of the model, as ITNs typically lose effectiveness over time and require replacement every 3 years due to the gradual decline in insecticidal effects. Accounting for this decay could affect the behaviour of the model and consequently its predictions. Furthermore, the availability of empirical country-specific data on treatment cases, improving the efficacy of ITN, creating income opportunities, and promoting social awareness programs would improve the robustness of the conclusions. These factors highlight the challenge of accurately determining the relative importance of each variable in real-world scenarios.\\
\textit{Social-learning} is a crucial behavioural component of the model. It plays a key role in how people accumulate information, understand beliefs and practices, and make decisions about the use of ITN within their communities. In addition to modern communication systems, social learning in many SSA countries occurs through local interactions and public health awareness campaigns, which vary notably from one country to another. Although we have assumed a common upper limit for all countries, having access to country-specific data would enhance the robustness of our results. A limitation of our behaviour-incidence model is the absence of data that accurately quantifies how ITN misuse might improve individuals' livelihoods or economic conditions; for example, estimates of bed nets used for fencing or fishing that improve the food security of the household. We had to rely on a calibrated range to estimate the relative benefit of improper ITN use. If such data were available, it would enable a more precise estimate of the subsidies required to improve daily productivity and encourage proper use of ITN.\\

In conclusion, understanding the dynamics of malaria prevalence and ITN usage in the context of GTS plans for SSA countries is crucial to developing new strategies \cite{bennett2017engaging} for malaria elimination. The objective of our analysis was to clarify the role of human behaviour in achieving the milestones of GTS 2025 and 2030 in 38 Sub-Saharan African countries. Despite exceptional efforts by national malaria programs, compromised ITN usage has presented challenges to public health initiatives and disease management. By fitting behaviour-incidence models to time series data on malaria incidence and ITN usage, we identified country-specific factors driving the success or failure of achieving the GTS goals for 2025 and 2030. Our findings suggest that a critical and often overlooked factor in malaria transmission is how individuals, families, and communities use ITNs as protective measures against infection.

\section*{Methods}

\subsection*{Data Availability}

\textbf{Data on malaria incidence}: The analysis utilized publicly available data from the \textbf{World Health Organization (WHO) Malaria Report 2021} (https://www.who.int/teams/global-malaria-programme/reports/world-malaria-report-2021), which provides detailed national-level malaria incidence cases per 1,000 individuals in 38 high-burden countries in Sub-Saharan Africa from 2000 to 2022. These data span trends over two decades, providing insight into malaria prevalence.

\textbf{ITN usage data}: Geospatial data on the proportion of the population using insecticide-treated nets (ITN) were sourced from the \textbf{Malaria Atlas Project (MAP)} (https://malariaatlas.org). This data set covers ITN usage (that is, the \textit{ proportion of the population who sleeps in an insecticide-treated net during a defined year}) over the same period (2000-2022), allowing for an in-depth analysis of the coverage and adoption rates of this critical preventive intervention.

Both data sets are publicly accessible via the links provided above and are integral to the temporal analysis presented in this study. Further details on data processing and analysis are included in the Supplementary Information.

\subsection*{Statistical Methods}
\textit{Pearson Correlation Coefficient:} We calculate the Pearson correlation coefficient between malaria incidence cases and ITN use from 2000-2020 for the 38 highest burden SSA countries (see Section S2.1 in the Supplementary Information). \\

\noindent\textit{Granger Causality:}
We conduct leave-one-out  Granger causality tests for the malaria incidence cases variables and ITN use of a fully specified vector autoregression model (VAR) (represented by a var model object), to assess whether each variable in a 2-D VAR model Granger causes another variable or not, as described in the S2.2 section of Supplementary Information.\\
\newline
\noindent\textit{Transfer Entropy:} We estimated the transfer entropy using the Binning method according to a uniform estimation approach between malaria incidence cases and ITN use. For more details on TE, see the S2.3 section of Supplementary Information.

\subsection*{The behaviour-Incidence Model}
The full set of equations and the descriptions of all parameters are given in Section S1 of the Supplementary Information. 

\subsection*{Estimation and Projection Methods}
In this section, we briefly describe the methodology to estimate the parameters of the behaviour incidence model and project country-specific strategic interventions related to the use of ITN.  \\
 \newline
\textit{Assumptions and Pre-estimation Calibration}

We simulate the behaviour-incidence model over 80 years to obtain the country-specific incidence pattern at the year 2000. During this calibration phase of the pre-estimation, we consider the dynamic model of malaria (excluding the behavioural component $x_h$ in the model S1) with specific baseline parameters to eliminate the transient phase of the model trajectories. The description and baseline values of the parameters for the behaviour-incidence model S8 are provided in Table S1 in the Supplementary Information. We assume that certain epidemiological parameters, such as mosquito biting rate ($\beta$), human recruitment rate ($\Lambda_h$), mosquito recruitment rate ($\Lambda_v$), and disease-induced mortality rate ($\delta_h$), vary between different countries due to differences in demographics, housing infrastructure, climate conditions, and other factors. To account for these variations, we calibrate these four parameters to align the behaviour-incidence model’s predicted malaria infection cases with the reported data for the year 2000. Additionally, we also ensure that the model’s predicted disease mortality rate (per 1,000) matches the available data for all 38 countries. This calibration process trains our behaviour-incidence model to fit country-specific systems. The calibrated parameter values are provided in Table S2 in the Supplementary Information.\\
 \newline
 \textit{Parameter Estimation}
 
In Phase-0 (starting from 2000 and until 2020), we use the full behaviour-incidence model (S8) to estimate three parameters: the mosquito biting rate ($\beta$), the social learning rate ($\kappa$), and the relative benefit of ITN misuse ($r$). Along with the behaviour-incidence model (S8), we also use an observational model based on the Poisson distribution to fit the reported malaria incidence and ITN usage data (a detailed discussion is given in Section S3 in the Supplementary Information). The observational model includes the parameter `reporting probability' which is estimated along with initial conditions. To calculate the  confidence intervals for the estimated parameters, we also use the bootstrap method, as described in  section S6 in the Supplementary Information .\\
\newline
\textit{Optimization \& Projection}

In Phase-I (start with 2020 until 2025), we project the behaviour-incidence model to 2025 based on optimizing three key parameters for each country: $\kappa$, $r$, and $b_0$. The parameter $\kappa$ serves as a proxy for social awareness and knowledge regarding proper ITN usage while $r$ reflects the socioeconomic conditions of the population. $b_0$ denotes ITNs efficacy. These three parameters represent distinct public health strategies aimed at enhancing the country-specific malaria elimination processes. If the country does not reach the 2025 milestone with fitted values obtained in Phase 0, we optimize the country-specific behaviour-incidence model to determine the optimal behavioural parameters and the ITN efficacy necessary to minimize the costs associated with malaria infection and its control. This process is repeated in Phase II (start with 2025 until 2030) for achieving the 2030 milestone. Detailed descriptions of this procedure are provided in Section S3 in the Supplementary Information.\\
\newline
\textit{K-Means} Clustering

We apply \textit{Kaiser criterion}, \textit{Factor analysis}, \textit{Elbow method,} and \textit{K-mean} clustering to identify underlying factors that explain the variance in the country-specific parameter regimes, aiming to better understand and predict strategic interventions related to ITNs usage. Factor analysis helps to identify correlations between the original variables and the factors. Variables with large loading values on a factor are considered highly influenced by that factor. The goal is to identify the common factors and their relationships with
observed data. The Kaiser rule is used in factor analysis to determine the components or factors, while the Elbow method helps identify the optimal number of clusters. \textit{K-mean clustering} groups countries into clusters based on significant variables and the optimal number of clusters that represent groups with distinct characteristics.  This approach allows us to group countries with similar epidemiological and behavioural factors, aiding in the development of targeted interventions.

\bibliography{main}


\section*{Author contributions statement}

L. conceived, conducted formal analysis and simulation, and wrote the first draft; T.O. conducted formal analysis and writing; M.T. conducted formal analysis and writing; I.D. conducted formal analysis, visualization and writing S.B. conceptualization,  formal analysis, simulation, validation, supervision, writing, and review. All authors
reviewed the manuscript. 

\section*{Additional information}
Codes are available upon request from the corresponding author.
 
\section*{Competing interests}
The authors declare no competing interests.

\section*{Declaration of use of AI tools}
The authors declare no usage of AI tool during the preparation of the manuscript.

\clearpage

\includepdf[pages=-]{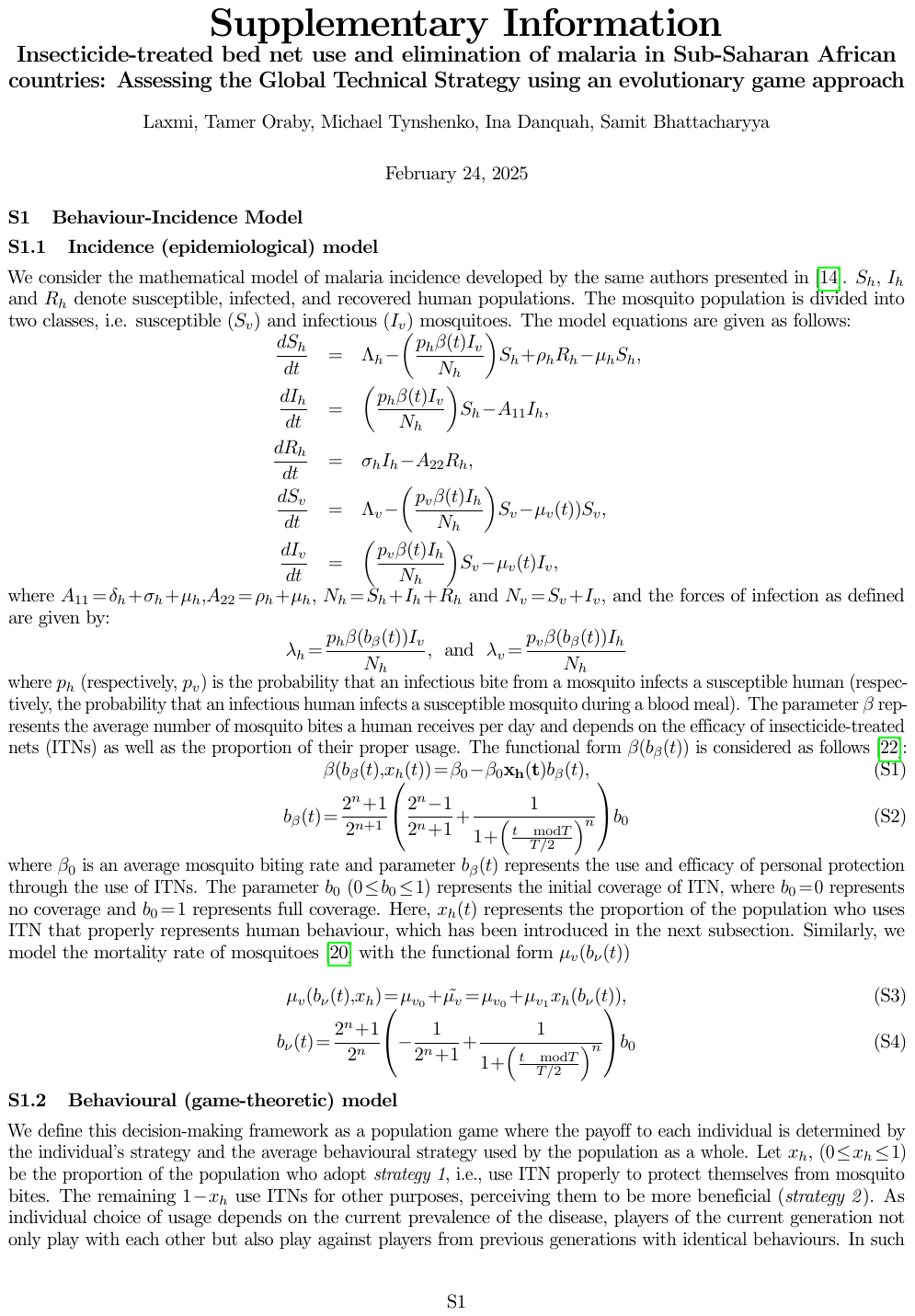} 

\end{document}


\maketitle

\section{Behaviour-Incidence Model}
\subsection{Incidence (epidemiological) model}
We consider the mathematical model of malaria incidence developed by the same authors presented in \cite{laxmi2022evolutionary}.  $S_h$, $I_h$ and $R_h$ denote susceptible, infected, and recovered human populations. The mosquito population is divided into two classes, i.e. susceptible ($S_v$) and infectious ($I_v$) mosquitoes. The model equations are given as follows:
\begin{eqnarray}
\label{freq:Epi}
\nonumber
\frac{dS_h}{dt} &=& \Lambda_h - \bigg(\frac{p_{h}\beta(t) I_v}{N_h}\bigg)S_h+\rho_h R_h - \mu_hS_h,\\
\nonumber
\frac{dI_h}{dt} &=&\bigg(\frac{p_{h}\beta(t) I_v}{N_h}\bigg)S_h - A_{11}I_h,\\
\nonumber
\frac{dR_h}{dt} &=& \sigma_h I_h - A_{22}R_h,\\
\nonumber
\frac{dS_v}{dt} &=& \Lambda_v -\bigg(\frac{p_{v}\beta(t) I_h}{N_h}\bigg)S_v-\mu_v(t))S_v,\\
\nonumber
\frac{dI_v}{dt} &=&\bigg(\frac{p_{v}\beta(t) I_h}{N_h}\bigg)S_v - \mu_v(t)I_v,
\end{eqnarray}
where $A_{11} = \delta_h+\sigma_h+\mu_h, A_{22} = \rho_h+\mu_h$, $N_h = S_h + I_h + R_h$ and $N_v = S_v + I_v$, and the forces of infection as defined are given by:
$$\lambda_{h} = \frac{p_h\beta(b_{\beta}(t)) I_v}{N_h}, ~~\mbox{and} ~~\lambda_{v} = \frac{p_{v}\beta(b_{\beta}(t)) I_h}{N_h}$$
where $p_{h}$ (respectively, $p_{v}$) is the probability that an infectious bite from a mosquito infects a susceptible human (respectively, the probability that an infectious human infects a susceptible mosquito during a blood meal). The parameter $\beta$ represents the average number of mosquito bites a human receives per day and depends on the efficacy of insecticide-treated nets (ITNs) as well as the proportion of their proper usage. The functional form $\beta(b_{\beta}(t))$ is considered as follows \cite{ngonghala2012periodic}:
\begin{equation}
\label{eqn: hello1}
\beta(b_{\beta}(t),x_h (t)) = \beta_0- \beta_0 {\bf{x_h (t)} }b_\beta(t),
\end{equation}
\begin{equation}
    b_{\beta}(t) = \frac{2^{n}+1}{2^{n+1}}\left(\frac{2^n-1}{2^n + 1} + \frac{1}{1 + \left(\frac{t\mod T}{T/2}\right)^n}\right)b_0
  \end{equation}
where $\beta_0$  is an average mosquito biting rate and parameter $b_{\beta}(t)$ represents the use and efficacy of personal protection through the use of ITNs. The parameter $b_0$  ($0 \le b_0 \le 1$) represents the initial coverage of ITN, where $b_0 = 0$ represents no coverage and $b_0 = 1$ represents full coverage. Here, $x_h(t)$ represents the proportion of the population who uses ITN that properly represents human behaviour, which has been introduced in the next subsection. Similarly, we model the mortality rate of mosquitoes \cite{ngonghala2014quantifying} with the functional form $\mu_{v}(b_\nu(t))$

\begin{equation}
\label{eqn: hello1}
\mu_{v}(b_{\nu}(t),x_h) = \mu_{v_0}+\Tilde{\mu_{v}}=\mu_{v_0}+\mu_{v_1}x_h (b_{\nu}(t)),
\end{equation}
\begin{equation}
    b_{\nu}(t) = \frac{2^{n}+1}{2^{n}}\left(-\frac{1}{2^n + 1} + \frac{1}{1 + \left(\frac{t\mod T}{T/2}\right)^n}\right)b_0
  \end{equation}
  
\subsection{Behavioural (game-theoretic) model}
We define this decision-making framework as a population game where the payoff to each individual is determined by the individual's strategy and the average behavioural strategy used by the population as a whole. Let $x_h, ~ (0 \leq x_h \leq 1)$ be the proportion of the population who adopt \textit{strategy 1}, i.e., use ITN properly to protect themselves from mosquito bites. The remaining $1 - x_h$ use ITNs for other purposes, perceiving them to be more beneficial (\textit{strategy 2}). As individual choice of usage depends on the current prevalence of the disease, players of the current generation not only play with each other but also play against players from previous generations with identical behaviours. In such strategic decision-making and social interaction, we assume that individuals emulate others' activities. Specifically, they sample other members at a constant rate, and if the payoff of the sampled person is higher, then the sampled strategy is adopted with a probability that is proportional to the expected gain in payoff \cite{bauch2005imitation, bhattacharyya2010game}. Individuals are assumed to switch between the two strategies depending on the perceived benefits either from using ITNs properly or improperly.\\
In game theory, individuals act rationally in choosing the strategy that results in a higher payoff. After an individual has access to an ITN, the individual selects the most preferred strategy that maximizes the associated expected utility. This adaptive behaviour is influenced by several social and economic factors. Let $L\in [0,\infty)$ denote the baseline daily productivity of an individual. Then improper use of ITNs, e.g., for fishing or agriculture, may increase baseline daily productivity. Let $r_L > 1$ be the proportional increment in the daily productivity. Then the perceived payoff for improper ITN-use is given by
\begin{equation}
f_{im} = L(r_L + 1).
\end{equation}
\noindent
In contrast, proper use of ITNs reduces the risk of infection through mosquito bites and depends on the proportions of current disease prevalence ($I_h$) and mosquito density ($N_v$) in the community. If $r_i$ is the risk of infection, then the expected payoff for proper use of ITNs is given by
\begin{equation}
f_{p} = r_i[1 -  b_{\beta}(t)](w_1 I_h + w_2 N_v),\label{UtilityFunction}
\end{equation}
where $w_1$ and $w_2$ are proportionality constants. It should be noted that perceived reward is also a function of the efficacy of ITN since people are aware of the efficacy of ITNs from the start. Thus, payoff gain for switching to the strategy of proper use of ITN is given by
$\Delta G =  f_p - f_{im}$
and accordingly, the evolution equation of $x_h$ (when $\Delta G > 0$) is given by:
\begin{eqnarray}
\label{freq:properUse}
\nonumber
\dot{x}_h &=& \varrho x_h (1 - x_h). \varsigma\Delta G,\\
{} &=& \kappa x_h (1 - x_h)\{-r + [1 - b_{\beta}(t)](w_1 I_h + w_2 N_v)\},
\end{eqnarray}
\noindent
where $k= \varrho \varsigma  r _i$ is the scaled emulation or imitation rate, and $r = \frac{L(r_L + 1)}{r_i}$ denotes the relative profit of improper usage of ITNs. This equation is similar to the replicator equation in population game \cite{hofbauer1998evolutionary}. It should be noted that the fraction of individuals $(1- x_h)$, who engage in improper use of ITN, satisfies the same equation (\ref{freq:properUse}).

\subsection{Integrated Behavioural-Incidence Model}
\begin{figure}[H]
    \centering
    \includegraphics[width=0.8\textwidth]{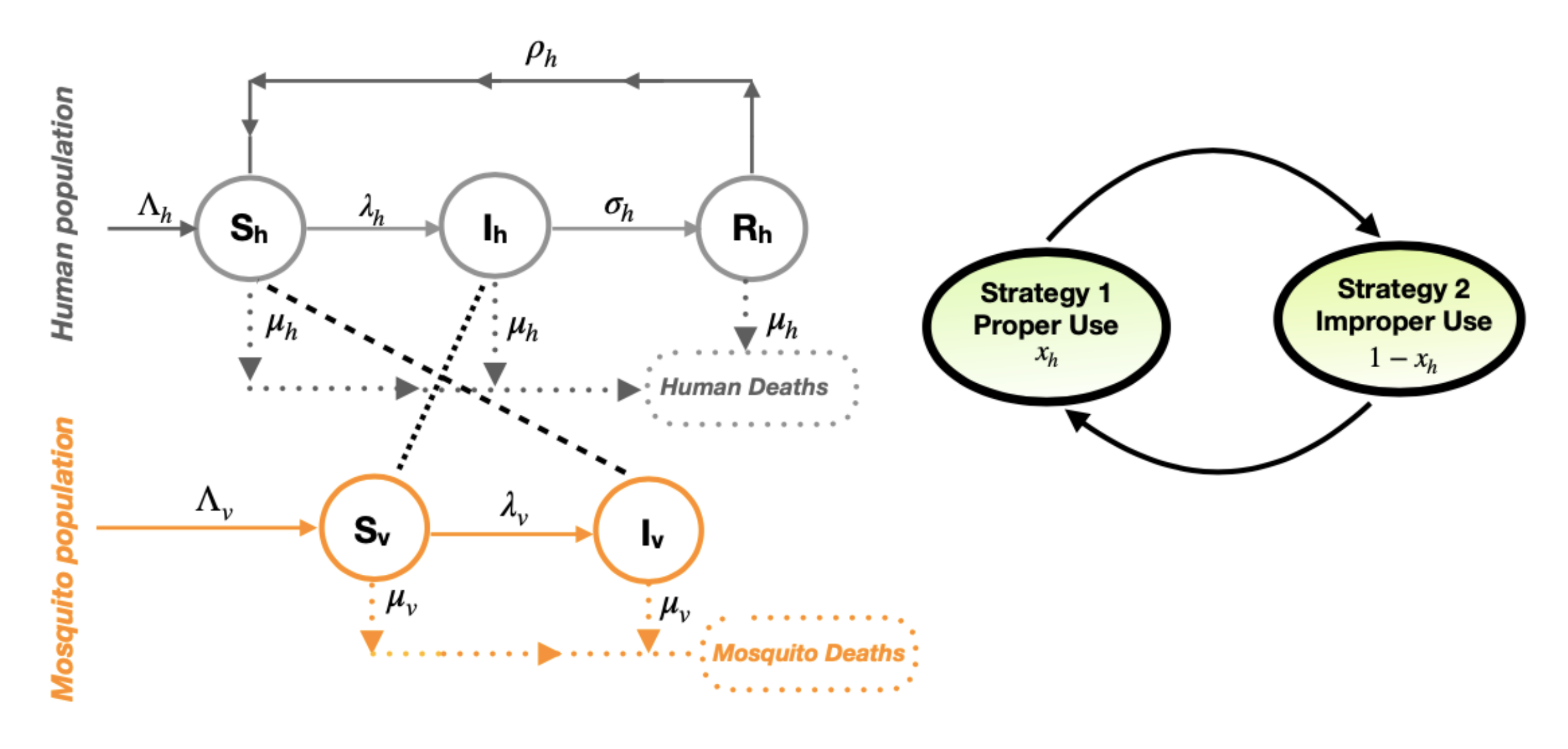}
   \vspace{-0.1cm}
\caption{\textbf{Schematic representation of the disease model \eqref{freq:Epi}} depicting transitions of humans and mosquitoes between different compartments (solid lines), the transmission of malaria from infectious mosquitoes to susceptible humans (dashed line), and the transmission of malaria from infectious humans to mosquitoes (dashed-dotted line). Moralities are denoted by dotted lines. Switching between strategies of ITN usages defines a feedback system—a part of the behaviour-incidence model (Right panel of the figure).}
    \label{fig:schematic ITN}
\end{figure}

We now integrate the incidence model given by Eqs. \eqref{freq:Epi} and the behaviour model given by \eqref{freq:properUse} into the coupled framework:
\begin{eqnarray}
\label{Model}
\nonumber
\dot{S}_h &=& \Lambda_h - \lambda_{h}S_h+{\rho_h}R_h - \mu_hS_h,
\nonumber\\
\dot{I}_h &=&\lambda_{h}S_h - A_{11}I_h,
\nonumber\\
\dot{R}_h &=& \sigma_h I_h - A_{22}R_h,\\
\dot{S}_v &=& \Lambda_v -\lambda_{v}S_v-\mu_v(t)S_v,\nonumber\\
\dot{I}_v &=&\lambda_{v}S_v - \mu_v(t)I_v,\nonumber\\
\dot{x}_h &=& \kappa x_h (1-x_h)\{-r+[1-b_{\beta}(t)](w_1 I_h+w_2 N_v)\}.\nonumber
\end{eqnarray}
  This model (\ref{Model}) is analyzed and use to validate the time series data of malaria incidence cases and ITN use. Figure \ref{fig:schematic ITN} represents the schematic of the behaviour-incidence model \ref{Model}. Throughout the analysis in this paper, we assume $b_\beta(t)=b_\nu(t)=b_0$, the constant efficacy of ITNs.
    \vspace{-0.5cm}

\section{Statistical Methods}
\subsection{Pearson Correlation Coefficient}
\label{PCC}
The correlation coefficient of two variables is a measure of their linear dependence. If each variable has $N$ observations, then the Pearson correlation coefficient is defined as
\begin{eqnarray}
  r(A,B)=\frac{1}{N-1}\sum_{i=1}^{N}\left(\frac{A_i-\overline{A}}{s_A}\right)\left(\frac{B_i-\overline{B}}{s_B}\right)  
\end{eqnarray}

where   $\overline{A}$   and $s_A$      are the mean and standard deviation of A, respectively, and  $\overline{B}$       and   $s_B$      are the mean and standard deviation of $B$. Once can also define the correlation coefficient in terms of the covariance of  $A$ and $B$:
\begin{eqnarray}
    r(A,B)=\frac{cov(A,B)}{s_A \, s_B}
\end{eqnarray}

\begin{figure}[H]
    \centering
    \includegraphics[width=1\textwidth]{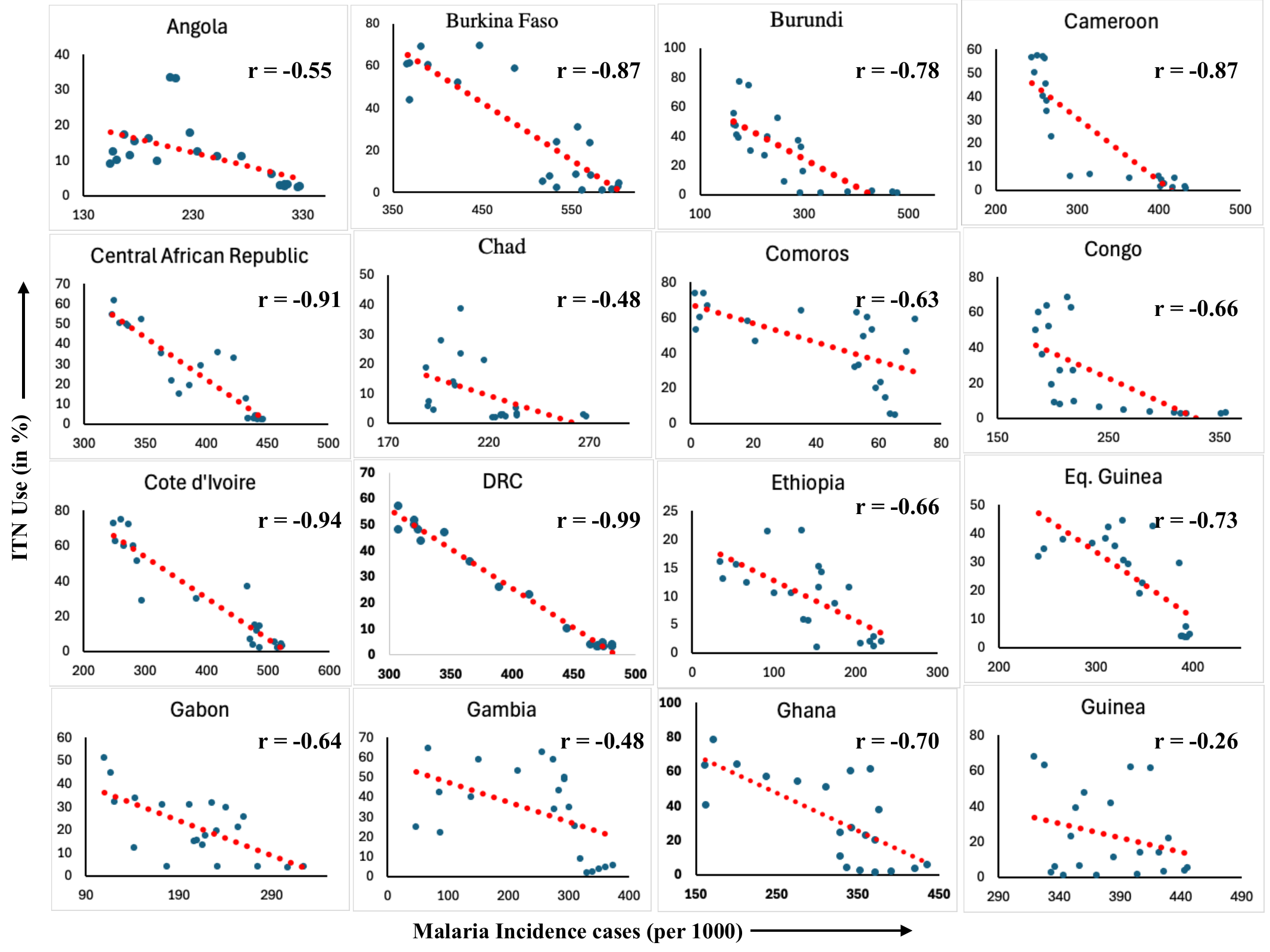}
     \caption{Scatter plots demonstrating the negative correlation between ITN use and Malaria incidence in mentioned 16 SSA countries}
    \label{fig:regressionplots1}
  \end{figure}
  
  \begin{figure}[H]
    \centering  \includegraphics[width=1\textwidth]{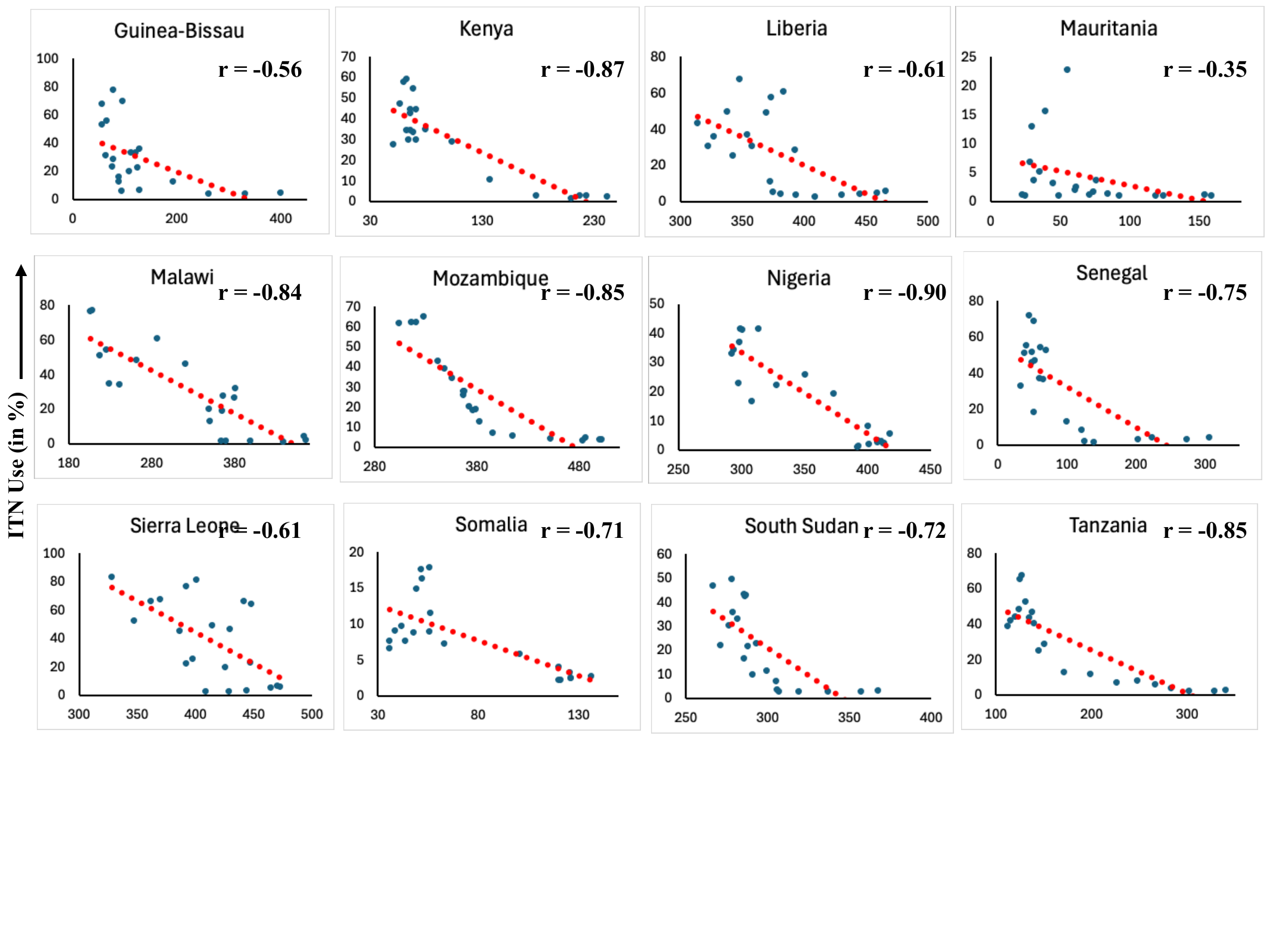}
    \vspace{-42mm} 

    \hspace{0.3cm}
     \includegraphics[width=1\textwidth]{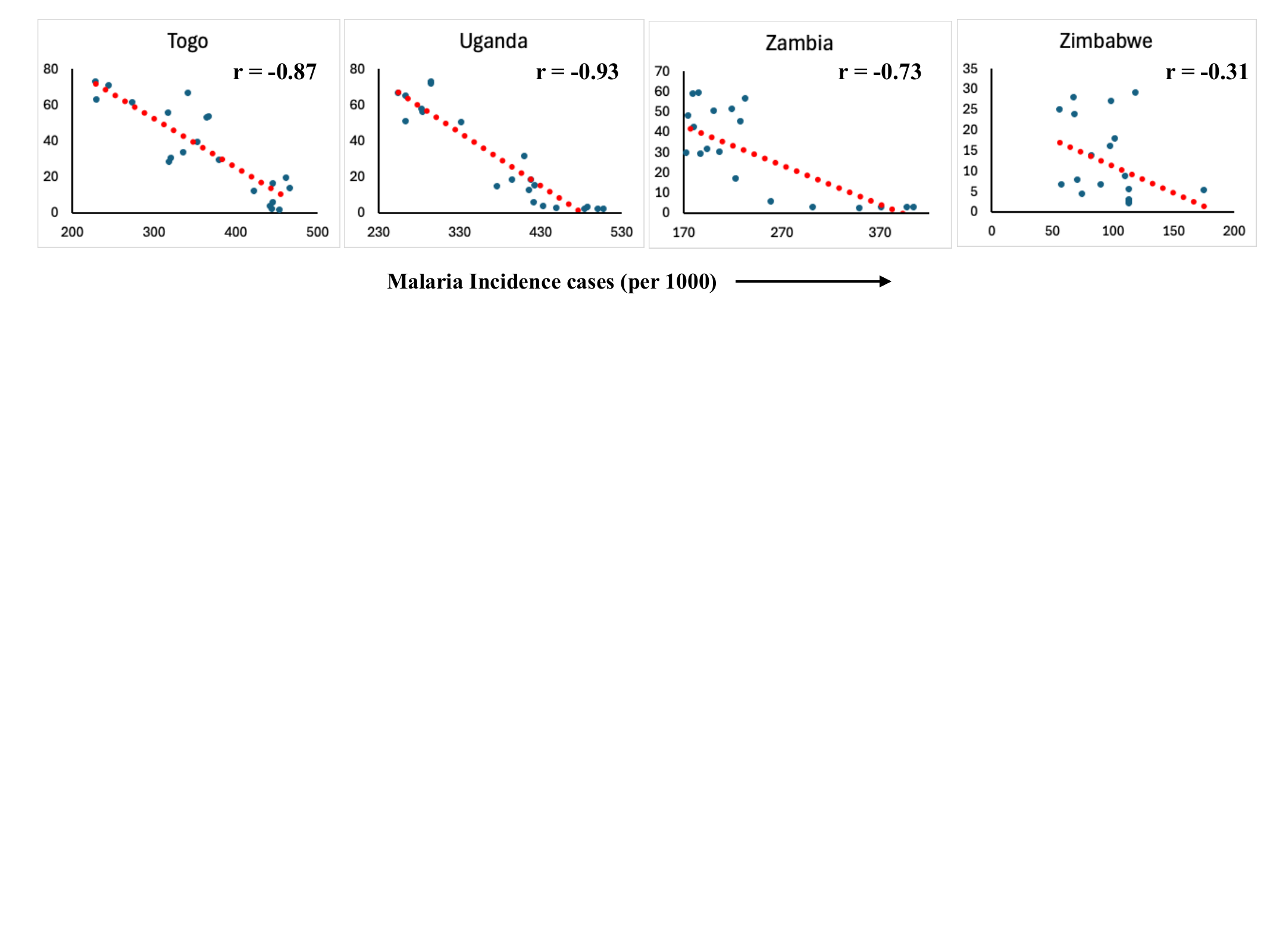}
    
   \vspace{-80mm}
    \caption{Scatter plots demonstrating the negative correlation between ITN use and Malaria incidence in mentioned 16 SSA countries}
    \label{fig:regressionplots2}
\end{figure}

 \subsection{Granger Causality}
\label{GC}
The Granger causality test is a statistical hypothesis test for determining whether one time series is useful in forecasting another, first proposed in 1969. Ordinarily, regressions reflect "mere" correlations, but Clive Granger argued that causality could be tested by measuring the ability to predict the future values of a time series using prior values of another time series. 
A time series $X$ is said to Granger-cause $Y$ if it can be shown, usually through a series of t-tests and F-tests on lagged values of $X$ (and with lagged values of $Y$ also included), that those $X$ values provide statistically significant information about future values of $Y$.

To implement a test of Granger causality, we perform the autoregressive specification of a bivariate vector autoregression. Assume a particular autoregressive lag length $p$, and estimate the following unrestricted equation by ordinary least squares (OLS):
\begin{eqnarray}
    y_t=c_1+\sum_{i=1}^{p}\alpha_i y_{t-i}+\sum_{i=1}^{p}\beta_i x_{t-i}+u_t
\end{eqnarray}
\begin{eqnarray}
    H_0:\beta_1=\beta_2 .....=\beta_p =0
\end{eqnarray}

Conduct an $F$-test of the null hypothesis by estimating the following restricted equation also by OLS:
\begin{eqnarray}
    y_t=c_1+ \sum_{i=1}^{p}\gamma_i y_{t-i}+e_t
\end{eqnarray}
Compare their respective sum of squared residuals $RSS_1=\sum_{t=1}^{T}\hat{u}_t^2 $, 
$RSS_0=\sum_{t=1}^{T}\hat{e}_t^2$.
If the test statistic 
$S_1=\frac{(RSS_0-RSS_1)/p}{RRS_1/(T-2p-1)} \sim  F_{p,T-2p-1}$,
is greater than the specified critical value, then reject the null hypothesis that $X$ is said to be \textbf{Granger-cause} $Y$. 

\subsection{Transfer Entropy}\label{TE}
Transfer entropy is rooted in information theory and based on the concept of Shannon entropy as a measure of uncertainty.
In 1948, Shannon introduced the concept of entropy (later referred to as Shannon entropy), to quantify the information  for a discrete random variable J with probability distribution, calculated as
\begin{equation}
     H_j=\sum_{j} p_j log_2 \frac{1}{p_j} =-\sum_{j} p_j log_2 {p_j}
\end{equation}
   To measure the information flow between two time series, Shannon entropy is coupled with the concept of Kullback–Leibler distance under the assumption that the underlying processes are Markov processes. 
Let $I$ and $J$  denote two discrete random variables with marginal probability distributions  $p(i)$ and $p(j)$, joint probability $p(i,j)$, whose dynamical structures correspond to a stationary Markov process of order $k$ (process $I$) and $l$ (process $J$). The average amount of measure of information  in $t+1$  observations of process $I$, once the previous    $k$  values are known, is given by
\begin{equation}
    h_I(k)=-\sum_{i} p(i_{t+1},{i_{t}}^{(k)}) log_2 p(i_{t+1}|{i_{t}}^{(k)})
\end{equation}
In the bivariate case information flow from process J to process I is measured by quantifying the deviation from the generalized Markov property   $ p(i_{t+1}|{i_{t}}^{(k)})= p(i_{t+1}|{i_{t}}^{(k)}, {j_{t}}^{(l)})$,  relying on the Kullback–Leibler distance. The formula for Transfer entropy  is given by 
\begin{equation}
    T_{J->I}(k,l)=\sum_{i} p(i_{t+1},{i_{t}}^{(k)},{j_{t}}^{(l)}) log_2 \frac{ p(i_{t+1}|{i_{t}}^{(k)},{j_{t}}^{(l)}))}{ p(i_{t+1}|{i_{t}}^{(k)})}
\end{equation}
where $T_{J->I}$ measures the information flow from $J$ to $I$.     $T_{I->J}$, as a measure for the information flow from $I $to $J$, can be derived analogously.

In this study, out of three different approaches (binning, nearest neighbour, linear), we discuss the binning estimator to evaluate the probability distribution function that constitutes the basis for TE in multivariate systems. In turn, each approach has to be paired with the choice of the time series' past values which contribute information to the knowledge of the present state of a given target time series. The first choice is the classical uniform embedding (UE) that considers a fixed amount of past terms for each series; the second approach is quite recent and employs a non-uniform embedding (NUE) \cite{faes2011information}, \cite{kugiumtzis2013direct} iteratively selecting the most informative terms through an optimization criterion.\\
We implemented the Binning estimator method in two different ways according to UE and NUE approaches in an organic toolbox MuTE in MATLAB \cite{montalto2014mute}, allowing straightforward comparisons between the methods.

\section{Baseline parameters, parameter estimation, likelihood function}
\label{estimation}
The values of the parameters (Table \ref{tab:Table 1} and Table \ref{tab:Table 2} respectively) and estimate 4 out of 14 epidemiological parameters that say transmission rate ($\beta$), recruitment rate of humans by natural birth ($\Lambda_h$), recruitment rate of mosquitoes by natural birth ($\Lambda_v$), and disease-induced human mortality rate ($\delta_h$) and 2 behavioral parameters that say imitation rate ($\kappa$) and relative cost of using bed nets properly($r$) and the reporting probability for all 38 SSA countries using time series data from the years 2000 to 2022.\\

Model parameters ($\beta, \kappa, r$) were estimated by minimizing the negative log-likelihood. We calculate the likelihood that a chosen parameter set $\Theta =(\beta,\kappa,r)$ explains the complete data  $Y=y(t_1), y(t_2),.....$ $y(t_n)$,
$n=21$  within the confines of the process and observation models. This likelihood function $L(\Theta)$ is a product of conditional likelihoods $L_{t_j}(\Theta)$, calculated  $t_j$ at each time for all 21 data points in time. We assume that the observed new infections (combined primary and secondary infections) each day follow a Poisson process with a mean of new infections predicted by the disease model. The log-likelihood function is defined as:
\begin{eqnarray}
    \log L(\Theta)=\sum_{j=1}^{n}\log L_{t_j}(\Theta),
\end{eqnarray}
where,
\begin{eqnarray}
    \log L_{t_j}(\Theta)=\log L_{t_j}(y({t_j})/\hat{y}_{\Theta}(t_j))=y({t_j})\log\hat{y}_{\Theta}(t_j)-\log(y({t_j})!)-\hat{y}_{\Theta}(t_j)
\end{eqnarray}
and, $\hat{y}_{\Theta}(t_j)$ is observed mean of new infections on day $t_j$ as integrated output from disease and observation models.\\

We estimate the initial conditions of the ODE system and simulate the integrated behaviour-incidence model exactly for 21 years to fit the time series of malaria incidence cases and the use of ITN to the variables $I_h$ and $x_h$ defined above in model(\ref{Model}). We estimate transmission ($\beta$), behavioural parameters $\kappa$, $r$, and the probability of reporting of all 38 countries as given in Table (\ref{tab:Table 1}).
As maximum likelihood estimates can be sensitive to the choice of initial values provided to the numerical optimization algorithm, we allow the solver in Matlab (version R2022a) to run till satisfy the good fitness criteria above $50\%$ $\&$ more to identify the parameters with greatest likelihood.\\

\section{Sensitivity Analysis}
Sensitivity analysis involves examining how the estimated parameters change with variations in the data. We use the Bootstrap method to determine the confidence interval for the estimator parameters. With estimated parameters $(\beta, \kappa,r)$, for a given time series of malaria incidence cases and ITN use from 2000-2020, create 50  new bootstrap samples by random sampling with replacement from the original datasets. For each new bootstrap sample, we perform MLE and obtain 50 new sets of bootstrap estimated parameters $(\beta_i, \kappa_i,r_i), i=1,2....n$. By using bootstrap samples, derive $95\%$ a confidence interval for each estimated parameter. This bootstrap-derived confidence interval helps assess the sensitivity of the parameter estimates to variations in the data, providing a robust measure for statistical inference.

\section{Optimization and Projection}
\label{Proj_opt}
In both Phase-I and Phase-II, the projection of malaria incidence cases and ITN usage extends from 2020 to 2025, and subsequently to 2030, for all 32 Sub-Saharan African countries. We excluded the six countries that showed a lackk of a good fit to the model. The initial step in this process involves evaluating whether the GTS 2025 milestones can be achieved using the same estimated parameters obtained in Phase-0. If the milestone is not met, we proceed by performing Latin Hypercube Sampling (LHS) to identify the control parameters ($\kappa$, $r$) or ($\kappa$, $r$, $b$), that minimize the associated costs defined in eqn \ref{eqn:GTScost} while achieving malaria control targets.\\
Following this, we assess whether the GTS 2030 milestones can be achieved with the optimized control parameters from Phase-I. If not, the same optimization procedure is repeated for Phase-II to determine the optimal strategies for the 2030 milestone.

\section{Bootstrap Method}
\label{BSM}
We use the Bootstrap method to determine the confidence interval for the estimator parameters. First, we create 50 new bootstrap samples by random sampling with replacements from the original datasets. For each new bootstrap sample, we perform MLE and obtain 50 new sets of bootstrap estimated parameters $(\beta_i, \kappa_i,r_i), i=1,2....n$. By using bootstrap samples, derive $95\%$ a confidence interval for each estimated estimated parameter. This bootstrap-derived confidence interval helps assess the sensitivity of the parameter estimates to variations in the data, providing a robust measure for statistical inference.

\newpage
\section*{Six Countries Showing Lack of Good Fit to the Behaviour-Incidence Model}
In our analysis, the integrated behaviour-incidence model failed to accurately capture the patterns of malaria incidence and insecticide-treated net (ITN) usage for six countries out of the 38 countries in Sub-Saharan Africa. Specifically, the model's predicted patterns of malaria incidence and ITN usage in \textit{Niger, Rwanda, Mali, Benin, Madagascar, and Eritrea} did not align with the empirical data. Figure 12 illustrates the discrepancies between the model's predictions and the observed patterns. In these six countries, the goodness-of-fit was below 20\% (except Benin and Mali), suggesting no consistent impact of ITNs on malaria incidence. This is further reflected in the correlation coefficients between malaria cases and ITN usage, which were positive for all six countries, contrary to the expected trend (Figure 13). See results for data analysis. \\

\textit{Niger} - The predicted malaria incidence and ITN usage in Niger do not align with the observed empirical patterns. The empirical data reveal a cyclic trend, with an average of 350 cases per 1,000 people. There was a decline in new cases, from 363.2 per 1,000 in 2000 to 299.3 in 2004, followed by a sharp rise from 307.5 in 2005 to 414.3 in 2010, with similar fluctuations continuing through 2020. In contrast, ITN usage follows a different trajectory, with almost no recorded use until 2005, and several declines in usage afterward, despite reaching 40\% by 2020. This inconsistency in patterns is driven by Niger's diverse climate, varying rainfall, unequal distribution of healthcare facilities, political conflicts, socioeconomic challenges, and uneven surveillance across regions \cite{tchole2024epidemiological}. For example, Niamey, Niger's rapidly growing capital, has high-quality medical care and accessibility, whereas Agadez, where most districts are located, suffers from poor infrastructure, limited healthcare access, terrorist attacks, and a significant influx of migrants. These factors may reduce the effectiveness of ITNs in controlling malaria incidence. Additionally, intermittent ITN usage spikes, driven by awareness campaigns, have occurred during this period. However, our behaviour-incidence model is based on simplified assumptions that do not account for these complexities, which may explain the poor fit between predictions and reality.\\
 
\textit{Benin} - Benin, a country in West Sub-Saharan Africa, exhibits a similar pattern to Niger. Despite high usage of insecticide-treated nets (ITNs), a major challenge contributing to the high malaria burden is the population's limited knowledge and awareness, leading to the trivialization of the disease and inadequate management of malaria control measures \cite{padonou2018knowledge}. Additionally, the characteristics of human habitats are not conducive to the effective use of indoor vector control interventions like ITNs. As highlighted in various studies, improving living conditions, alongside enhancing awareness and knowledge of disease management, is a crucial next step for malaria elimination efforts in Benin\cite{damien2023human}. \\
 
\textit{Mali}- Mali has apparently constant malaria incidence from 2000 to 2020 in spite of steady increases in ITNs usages up to 50\% from 2005. The major reason of high persistence of malaria in Mali is its geography. Malaria epidemiological factors in Mali are associated with the four eco-climatic zones in the country: (1) the Sahara zone, with a short rainy season where epidemics of malaria could occur; (2) the Sahel zone, with mostly irrigated rice production enhancement projects where malaria epidemiology varies according to water used and agricultural activities; (3) the Sudanese-Guinean zone, where the transmission can last up to 6 months a year; and, (4) the Inner Delta of the Niger River, where malaria transmission continues year-round \cite{toure2022trends}. Apart from these, low awareness and knowledge, lack of proper infrastructure, decaying ITN efficacy, and inconsistent replacement campaigns are the most important factors inhibiting the proper indoor usages of ITNs in Mali. These underlying elements may disrupt the expected impact of net usage, leading to periods of high malaria transmission despite a robust culture of ITN use. A model including such features may better explain the dynamics of the disease and ITN usage in this country.\\

 \textit{Rwanda} - In comparison with others, Rwanda exhibits a very interesting pattern from 2000 to 2020. It indicates very good malaria control programme since 2000 to 2012, with lowest number of reported cases around 50 per 1,000 population. Despite these malaria control efforts, Rwanda has experienced an eight-fold increase in reported malaria cases countrywide (in all 30 districts) between 2012 and 2016. During the same period, malaria incidence increased from 48 malaria cases per 1,000 population in 2012 to 403 malaria cases per 1,000 population at risk in 2016 \cite{karema2020history}. Several factors are hypothesized to have contributed to this increase. This included: climatic variations (i.e. increased temperature and precipitation), delay of LLIN supply and delivery, insecticide (pyrethroid) resistance, short-term (approximately 2 years) durability of LLINs, substandard LLINs (low insecticide content and physical deterioration), change of vector behaviours towards more outdoor biting, as well as failure to provide adequate funding for universal coverage of key malaria interventions on time \cite{monroe2022reflections}. Our behaviour-incidence model is based on simple assumptions and does not include such variability in population demographics, climate, and socioeconomic conditions. \\

\textit{Madagascar \& Eritrea - } In contrast to all other four countries, Madagascar and Eritrea exhibit very low incidences—around 20-50 cases per 1,000 population throughout the period 2000-2020, with an increasing trend of cases in Eritrea from 2010, despite consistent usages of ITNs from 2000. In Madagascar, self-efficacy to prevent malaria and self-efficacy to obtain bed nets were associated with higher reported net use \cite{storey2018associations}. However, lack of modern healthcare systems and infrastructure, low economic productivity are playing a major role behind the recent increase in malaria prevalence. In Madagascar, almost all individual production units in the country are informal, and it was observed an inverse relationship between an increased informal sector and economic growth in the country Its documented in \href{https://www.elibrary.imf.org/view/journals/002/2023/118/article-A002-en.xml}{IMF eLIBRARY } \cite{IMF} . With the right policies (such as promoting formalization, creating public-private partnerships, and investing in human capital), governments can gradually improve health infrastructure over time in Madagascar. \\

In addition to the six countries, \textit{Nigeria} also falls into cluster 5, characterized by a high mean value of relative benefit from improper ITN use and low usage rates (see Figure \ref{fig:k-cluster}). Despite Nigeria's high GDP, it remains one of the four African countries responsible for half of the global malaria mortality \cite{world2022world}. The 2013 Nigeria DHS report suggests that the large gap in ITN usage may be due to the survey being conducted during a low malaria transmission season, though independent studies from the mid-2010s also reported low net use. Other studies have identified reasons for low ITN usage, including heat, low mosquito activity, fear of chemicals, lack of space, difficulty in hanging nets, preference for alternative preventive measures, ignorance, and cultural beliefs. Interestingly, the higher a caregiver’s perceived severity of malaria or belief in the effectiveness of ITNs in preventing infection, the lower the reported bed net use \cite{storey2018associations}. Moreover, large parts of central and northeastern Nigeria face significant challenges in ITN distribution campaigns \cite{bertozzi2021maps}. This suggests that, despite Nigeria's relatively strong economy, economic strength alone does not guarantee effective ITN utilization. However, since our behaviour-incidence model is based on simplified assumptions, it may need further refinement to better capture these dynamics.

\begin{table}[H]
\caption{Descriptions, baseline values and references of parameters (Model equations \ref{Model})}
\vspace{2mm}
\begin{center}
\resizebox{\textwidth}{!}{
\begin{tabular}{ p{0.085\linewidth}p{0.35\linewidth}p{0.15\linewidth}p{0.2\linewidth}p{0.25\linewidth}}
\hline
\hline
Parameter & Description and dimension & Baseline Value  & Range & Reference \\
\hline
 \hline  
$\mu_h$& Human natural Death rate (Day$^{-1}$)& $1/(55 \times 365)$ & $[1/72,1/35] \times 1/365$&  \href{https://www.cia.gov/library/publications/the-world-factbook/rankorder/2102rank.html}{Central Intelligence Agency (CIA) (2014b)}\\

$\mu_{v_0}$ & Mosquito natural Death rate (Day$^{-1}$)& $1/14$ & $[1/21,1/14]$ & \cite{davidson1953field, gilles1993bruce}\\

$\mu_{v_1}$ & pyrethroids dependent death rate of mosquito & $1/14$&$[1/21,5/10]$&\cite{ngonghala2016interplay, lines1987experimental}\\

  $\rho_h $ & The rate humans loose malaria immunity (Day$^{-1}$) &$1/(5\times 365)$ &$[55/10^{6}, 11/ 10^{3}]$ & \cite{chitnis2008determining}\\
 
$\sigma_h$ & Rate at which infectious human acquire immunity (Day$^{-1}$)& $1/285$ & $[14/10^4, 17/10^3]$&\cite{molineaux1979assessment} \\

$p_{v}$  &  the probability that a bite to an infectious  human will infect a susceptible mosquito &0.48 &[0.072, 0.64]&\cite{ngonghala2016interplay}  \\

$\beta_0$ &  Mosquito biting rate per day & - &[1/10, 1]&\cite{molineaux1980garki, gupta1994theoretical}\\

$b_0$ & implementation and efficacy parameter  of personal protection by ITN's & - &[0, 1] & \cite{ngonghala2016interplay} \\

$\kappa$ & the imitation rate per day & - &[0.001, 0.2] & \cite{laxmi2022evolutionary}\\
  
$L$ & the daily baseline productivity of an individual   & 1 & & \cite{laxmi2022evolutionary}\\
 
$r$ & the related cost of using ITN properly  & 1 & [8/10,18/10] & \cite{laxmi2022evolutionary}\\
	   
$w_1$ & Sensitivity parameter to infected human& 1/100 & - & \cite{laxmi2022evolutionary}\\

$w_2$ & Sensitivity parameter to number of mosquitoes& 1/6000 &- &\cite{laxmi2022evolutionary}\\

$\eta_1$& Cost associated with infection & 0.2 & - &Calibrated\\

$\eta_2$& Cost associated with imitation rate & 0.1 &-  &Calibrated\\
$\eta_3$& Cost associated with relative cost& 0.2 &  -&Calibrated\\

$\eta_4$& Cost associated with efficacy of ITN's & 0.01 &-  &Calibrated\\
  \hline

\end{tabular}}
\label{tab:Table 1}
\end{center}
\end{table}

\begin{table}[H]

\begin{center}
\begin{tabular}{|p{2.4cm}|p{3cm}|p{3cm}|p{3cm}|}\hline
{Country name}
&{$\Lambda_h$}& {$\Lambda_v$}& {$\delta_h$}
\\\hline\hline
 \hline
Ethiopia & 89.6637609	 & 128571.4286	 & 0.001674	\\ 
\hline
Gambia & 71.73100872	&128571.4286	& 0.0135  \\
 \hline  
Ghana & 107.5965131 &	128571.4286	& 0.0144 \\
\hline
Mauritania & 89.6637609	& 128571.4286 &	0.0288\\
\hline
Togo & 91.45703611 &	128571.4286	& 0.01278	\\
\hline
Senegal & 73.52428394 &	218571.4286 &	0.0306 \\
  \hline
 Eq Guinea & 80.69738481 & 128571.4286 & 0.02016 \\
  \hline
Kenya & 71.73100872 &	205714.2857 & 0.01044  \\
\hline
Burkino Faso  & 179.3275218	& 128571.4286 & 0.0432 \\
\hline	
Gabon  &  71.73100872	 & 1157142.857 & 0.01026 \\
 
\hline
Malawi & 89.6637609 & 128571.4286 & 0.01782 \\
\hline

Mozambique & 107.5965131 &	128571.4286 & 0.0225   \\
\hline 
Sierra Leone & 103.113325	& 308571.4286 & 0.0396\\
\hline	  
Tanzania & 71.73100872 & 244285.7143 & 0.01494\\
 \hline
 Zambia & 80.69738481	& 771428.5714 & 0.0117 \\
  \hline	   
  Somalia  & 53.79825654 & 153231.4286 & 0.00774 \\
  \hline
 Cameroon & 89.6637609	& 128571.4286 & 0.0198 \\
  \hline
CAR & 80.69738481 & 128571.4286 & 0.01242 \\
  \hline
  Liberia & 71.73100872 & 900000 & 0.0243 \\
  \hline
  Zimbabwe &26.89912827	&141428.5714 & 0.010998 \\
  \hline
  Chad & 91.45703611	& 128571.4286 & 0.01656 \\
  \hline
  the Congo  &78.90410959 &	128571.4286	& 0.01386\\
  \hline
  Cote d'Ivoire & 125.5292653	& 128571.4286	& 0.02394 \\
  \hline
  DRC & 143.4620174 & 	128571.4286	& 0.0486\\
  \hline
  Gui-Bissau & 91.45703611	&411428.5714 &	0.01404\\
  \hline
  Nigeria & 98.63013699	& 257142.8571& 0.02106 \\
  \hline
  Uganda & 98.63013699	& 128571.4286 & 0.02142 \\
  \hline
  Angola & 67.24782067	& 514285.7143 & 0.02142 \\
  \hline
  Burundi
  & 89.6637609 & 771428.5714	& 0.02124 \\
  \hline
  the Comoros & 53.79825654	& 810000 & 0.666 \\
  \hline
\end{tabular}%
\end{center}
\caption{Country-specific parameter values obtained during pre-estimation calibration. $\Lambda_h$: Per capita human recruitment rate, $\Lambda_v$: per capita mosquito recruitment rate, $\delta$: per capita daily disease-induced mortality rate.}
\label{tab:Table 2}
\end{table}

\newpage

\begin{table}[H]
\tiny
\begin{center}
\begin{tabular}{|p{2cm}p{3.2cm}p{2.5cm}p{2.5cm}p{1cm}p{1cm}|}
\hline
Country & Transmission rate ($\beta$) & Imitation rate ($\kappa$)& Relative cost ($r$) & GOF$_1$&GOF$_2$
\\\hline
 \hline
Ethiopia & 0.6961\newline (0.690, 0.699) & 0.293 \newline (0.280, 0.316) & 0.0087  \newline (0.0081, 0.0012) &0.69& 0.51\\ 
\hline
   Gambia &  1.746 \newline (1.727, 1.748)  & 0.450  \newline (0.450, 0.460) & 0.009 \newline (0.008, 0.013)&0.86&0.60	\\
 \hline  
Ghana & 1.790 \newline (1.779, 1.790) &0.450\newline (0.434, 0.458) &	0.08 \newline (0.077, 0.087)	&0.63&0.84 \\
\hline
Mauritania & 0.899 \newline (0.899, 0.907) & 0.011 \newline (0.010, 0.012)	& 0.470 \newline (0.460, 0.484)&0.70& 0.50\\
\hline
Togo & 1.557 \newline (1.556, 1.560) &0.550 \newline (0.545, 0.554) &	0.100 \newline (0.090, 0.120) &0.73&0.84\\
\hline
Senegal & 0.816 \newline (0.815, 0.820)  & 0.82 \newline (0.819, 0.829) &	 0.18 \newline (0.176, 0.181) &0.54 & 0.54\\
  \hline
 Eq Guinea & 1.310 \newline (1.307,1.311) & 0.7 \newline (0.694, 0.705)  & 0.25 \newline (0.248, 0.2533) &0.66&0.65\\
  \hline
Kenya & 0.820 \newline (0.813, 0.827) & 0.805\newline (0.803, 0.808)  &	0.18 \newline (0.172, 0.180)& 0.89& 0.78 \\
\hline
Burkina Faso & 1.487 \newline (1.485, 1.491) & 0.753 \newline (0.750, 0.755)	& 0.293 \newline (0.289, 0.299) &0.70& 0.73 \\
\hline	
Gabon  & 1.011 \newline (1.009, 1.055)  & 0.55 \newline (0.541, 0.551) & 0.120 \newline (0.114, 0.131) &0.68 & 0.40\\
\hline
Malawi & 3.316 \newline (3.314, 3.318) &  0.60 \newline (0.596, 0.607) & 0.090 \newline (0.086, 0.092) & 0.86 & 0.80\\
\hline
Mozambique &  3.902 \newline (3.900, 3.903) & 0.4 \newline (0.393, 0.405) & 0.014 \newline (0.012, 0.016) & 0.80& 0.80  \\
\hline 
Sierra Leone & 1.338 \newline (1.337, 1.341) & 0.50 \newline (0.496, 0.507)	& 0.31 \newline (0.305, 0.311) & 0.71 & 0.80\\
\hline	  
Tanzania & 0.966 \newline (0.966, 0.970) &0.795 \newline (0.791, 0.802)	 &0.201 \newline (0.197, 0.207) & 0.75 & 0.64\\
 \hline
 Zambia & 1.148 \newline (1.143, 1.148) &0.50 \newline (0.582, 0.507) & 0.08 \newline (0.075, 0.081) & 0.72 & 0.69\\
  \hline	   
  Somalia  & 0.704 \newline (0.703, 0.705) & 0.679 \newline (0.672, 0.703)& 0.095 \newline (0.085, 0.099)& 0.51& 0.53\\
  \hline
 Cameroon & 1.906 \newline (1.906, 1.915) & 0.58 \newline (0.579, 0.599)& 0.11 \newline (0.105, 0.118) & 0.80 & 0.82\\
  \hline
CAR & 1.993 \newline (1.991, 1.994) & 0.5 \newline (0.487, 0.513)& 0.14 \newline (0.138, 0.144) & 0.76& 0.73\\
  \hline
  Liberia & 1.164  \newline (1.163, 1.166)& 0.208 \newline (0.205, 0.225)& 0.026 \newline (0.020, 0.028)& 0.67 &0.67\\
  \hline
  Zimbabwe & 0.708 \newline (0.707, 0.709)&	0.43 \newline (0.428, 0.434) & 0.07 \newline (0.068, 0.072)& 0.51 & 0.73\\
  \hline
  Chad & 6.173 \newline (6.170, 6.177) & 0.68 \newline (0.676, 0.682) & 0.17 \newline (0.160, 0.171) & 0.61 & 0.56\\
  \hline
  the Congo  & 1.370 \newline (1.370, 1.372) & 0.55 \newline (0.546, 0.551) & 0.12	\newline (0.113, 0.123) & 0.52 & 0.61 \\
  \hline
  Cote d'Ivoire & 2.228 \newline (2.228, 2.229) & 0.4 \newline (0.391, 0.400)	& 0.009 \newline (0.008, 0.009) & 0.80 & 0.78 \\
  \hline
  DRC & 3.235 \newline (3.233, 3.236)  & 0.35 \newline (0.336, 0.352) & 0.01  \newline (0.009, 0.012)& 0.80 & 0.83\\
  \hline
  Gui-Bissau & 0.858 \newline (0.858, 0.859) & 0.81 \newline (0.807, 0.818)	& 0.15 \newline (0.143, 0.154) & 0.53 & 0.50  \\
  \hline
  Nigeria & 0.725 \newline (0.725, 0.726) & 0.4 \newline (0.390, 0.409) & 0.26 \newline (0.252, 0.262) & 0.77 & 0.79 \\
  \hline
  Uganda & 2.472 \newline (2.472, 2.476) & 0.6 \newline (0.594, 0.604)& 0.1 \newline (0.092, 0.104) & 0.83 & 0.86\\
  \hline
  Angola & 1.455 \newline (1.453, 1.456) & 0.55 \newline (0.541, 0.553)	& 0.09 \newline (0.087, 0.096) & 0.77 & 0.66 \\
  \hline
  Burundi &  0.432 \newline (0.431, 0.432) & 0.379 \newline (0.369, 0.379)& 0.002 \newline (0.002, 0.003) & 0.81 & 0.81\\
  \hline
  the Comoros & 0.451 \newline (0.450, 0.452)& 0.451 \newline (0.428, 0.451) & 0.72 \newline (0.067, 0.073) &  0.57 & 0.50 \\
  \hline
    Guinea & 5.467 \newline (5.462, 5.471)	& 0.350 \newline (0.349, 0.355)& 0.008 \newline (0.007, 0.008) & 0.42 &0.43\\
  \hline
   South Sudan & 4.466 \newline (4.462, 4.470)	& 0.336 \newline (0.335, 0.337) & 0.117 \newline (0.117, 0.119)& 0.40 &0.63 \\
  \hline
\end{tabular}%
\end{center}
\caption{Estimated parameter values in Phase-0 for all 32 SSA countries. GOF$_1$ and GOF$_2$ indicate the goodness-of-fit for reported cases and ITN usages. The confidence intervals of each estimated value obtained from the bootstrap method are also given here.}
\label{tab:Table 3}
\end{table}

\begin{table}[H]
\begin{center}
\adjustbox{scale=1}{
\begin{tabular}{|p{2cm}p{1cm}p{1.2cm}p{1.2cm}p{1.2cm}p{3.1cm}p{2cm}|}
\hline
{Country Name}
&{Path} & { $\kappa$}& {$r$} & {$b$}&{focused Intervention (majorly)}& {Relative ITN increase}
\\\hline
 \hline
Ethiopia & $A{1}$ & 0.293 & 0.008 & - & No additional & 90.47\%\\ 
\hline
Gambia & $A{2}$ & 4.406 & 0.007 & -	& ITN campaigns  & $\ast$\\
 \hline  
Ghana & $A{2}$ & 4.500 &	0.0008 & -	& ITN campaigns  & 31.69\%\\
\hline
Mauritania  &$A{1}$ & 0.47 & 0.011 &  - & No additional & 250.94\% \\
\hline
Togo & $B{2}$  & 4.5854	& 0.02 & - &ITN campaigns  & 11.75\%\\
\hline
Senegal & $B{2}$  &  4.6226 & 0.043 & -& ITN campaigns  & 59.8\%	  \\
  \hline
 Eq. Guinea & $B{2}$  & 4.645	& 0.002 & - &ITN campaigns  & 276.17\%\\
  \hline
Kenya &  $B{2}$  &  4.1675 & 0.003 & -	& ITN campaigns & 138.89\% \\
\hline
Burkina Faso & $B{2}$  & 4.5748	& 0.006 & - & ITN campaigns & 19.14\%\\
\hline	
Gabon  & $B{3}$   & 1.406  & 0.061 & 0.5997 & ITN campaigns \&  ITN efficacy & 156.33\%  \\
\hline
Malawi & $B{3}$  & 4.839  & 0.0142  & 0.595 & ITN campaigns \&  ITN efficacy & 52.98\%  \\
\hline
Mozambique & $B{3}$  & 4.983	& 0.009 & 0.599 & ITN campaigns \&  ITN efficacy & 16.21\% \\
\hline 
Sierra Leone & $B{3}$ &  4.533	& 0.0145 & 0.576 & ITN campaigns \&  ITN efficacy & 46.64\% \\
\hline	  
Tanzania & $B{2}$  & 4.239	 & 0.009 &  - & ITN campaigns  & 161.33\%\\
 \hline
 Zambia & $B{3}$  & 1.6354  & 0.0735 & 0.5972 & ITN campaigns \&  ITN efficacy & 59.41\% \\
  \hline	   
  Somalia  & $B{2}$  & 4.692 & 0.003 & - & ITN campaigns  & 192.04\%\\
  \hline
 Cameroon &  $B{3}$ & 3.7747  & 0.004 & 0.586 &ITN campaigns \&  ITN efficacy & 47.48\%\\
  \hline
CAR &  $B{2}$ & 4.822  & 0.009 & - & ITN campaigns & 114.95\%  \\
  \hline
  Liberia &  $B{3}$ & 4.6802 & 0.0146  & 0.6928 & ITN campaigns \&  ITN efficacy & $\ast\ast$ \\
  \hline
  Zimbabwe & $B{2}$  & 4.6089	&  0.0225  & - & ITN campaigns & 88.49\% \\
  \hline
  Chad & $B{3}$ & 3.222  & 0.058 & 0.597 & ITN campaigns \&  ITN efficacy & 77.75\%\\
  \hline
  The Congo  &  $B{2}$  &  4.3830 & 0.0068  & - &ITN campaigns & 20.36\%	 \\
  \hline
  Cote d'Ivoire & $B{3}$  &  4.7327 & 0.0041  &  0.5848	& ITN campaigns \&  ITN efficacy & 50.77\% \\
  \hline
  DRC & $B{3}$  &  3.5274	& 0.0057 & 0.585 & ITN campaigns \&  ITN efficacy & 52.53\%\\
  \hline
  Gui-Bissau & $B{3}$ &  1.5059 	& 0.129 & 0.591 & ITN campaigns \&  ITN efficacy & 5.12\%   \\
  \hline
  Nigeria & $B{2}$ & 4.777 & 0.013 & - & ITN campaigns & 115.57\% \\
  \hline
  Uganda & $B{3}$  &  3.8454	& 0.0195 & 0.598  & ITN campaigns \&  ITN efficacy & 77.63\%\\
  \hline
  Angola & $B{3}$ & 0.7123 & 0.063 & 0.596 & ITN campaigns \&  ITN efficacy & $\ast\ast$ \\
  \hline
  Burundi &  $B{3}$  & 1.6944 & 0.00038 & 0.593 & ITN campaigns \&  ITN efficacy & 1.99\% \\
  \hline
  Comoros & $B{3}$ & 0.4993  & 0.071 & 0.581 & ITN campaigns \&  ITN efficacy & 50.75\% \\
  \hline
  Guinea & $B{3}$	&  3.475 & 0.0024 & 0.698 & ITN campaigns \&  ITN efficacy &  $\ast\ast$ \\
  \hline
  South Sudan &  $B{2}$ 	& 4.635 & 0.0471 & - & ITN campaigns & 148.54\% \\
  \hline
\end{tabular}}%
\label{tab:Table 4}
\vspace{0.5cm}
\caption{Optimal path and parameter values, targeted intervention, and relative increase of ITN usages (concerning 2022 as a baseline value) for achieving the 2025 milestone (Phase-I). $\ast$ indicates there is no change in ITN usages, and $\ast\ast$ indicates no change or even less coverage of ITN usage with higher bed net efficacy. Details are discussed in the Conclusion section in the main text.}
\end{center}

\end{table}

\begin{table}[H]
\begin{center}
\adjustbox{scale=1}{
\begin{tabular}{|p{2cm}p{1cm}p{1.2cm}p{1.2cm}p{1.2cm}p{3.1cm}p{2cm}|}
\hline
{Country Name}
&{Path} & { $\kappa$}& {$r$} & {$b$}&{focused Intervention (majorly)}& {Relative ITN increase}
\\\hline
 \hline
Ethiopia & $A{11}$ & 0.293 & 0.008 & - & No additional & 144.85\%\\ 
\hline
Gambia & $A{21}$ & 4.406 & 0.007 & -	& No additional  & $\ast$\\
 \hline  
Ghana & $A{21}$ & 4.500 &	0.0008 & -	& No additional  & $\ast$\\
\hline
Mauritania  &$A{11}$ & 0.47 & 0.011 &  - & No additional & 396.98\% \\
\hline
Togo & $B{23}$  & 4.847	& 	0.0174	 & 0.59 & ITN efficacy  & 13.20\%\\
\hline
Senegal & $B{21}$  &  4.6226	 & 0.0432 & -& No additional & 64.8\%	  \\
  \hline
 Eq. Guinea & $B{21}$  & 4.645	& 0.002 & - & No additional  & 280.80\%\\
  \hline
Kenya &  $B{21}$  &  4.1675 & 0.003 & -	& No additional & 143.84\% \\
\hline
Burkino Faso & $B{21}$  & 4.5748	& 0.006 & - & No additional& 21.50\%\\
\hline	
Gabon  & $B{31}$   & 1.406  & 0.061 & 0.5997 & No additional &  $\ast$  \\
\hline
Malawi & $B{3}$  & 4.839  & 0.0142  & 0.595 & No additional & 54.78\%  \\
\hline
Mozambique & $B{32}$  & 4.993		& 0.006	& 0.69 &   ITN efficacy & $\ast$  \\
\hline 
Sierra Leone & $B{31}$ &  4.533	& 0.0145 & 0.576 & No additional & $\ast$ \\
\hline	  
Tanzania & $B{21}$  & 4.239	 & 0.009  &  - & No additional & $\ast$\\
 \hline
 Zambia & $B{31}$  & 1.6354  & 0.0735 & 0.5972 & No additional & $\ast$ \\
  \hline	   
  Somalia  & $B{21}$  & 4.692 & 0.003 & - & No additional & 221.96\%\\
  \hline
 Cameroon &  $B{31}$ & 3.7747  & 0.004 & 0.586 &No additional & 53.02\%\\
  \hline
CAR &  $B{21}$ & 4.822  & 0.009 & - & No additional & 118.49\%  \\
  \hline
  Liberia &  $B{31}$ & 4.6802 & 0.0146  & 0.6928 & No additional & $\ast\ast$ \\
  \hline
  Zimbabwe & $B{21}$  & 4.6089	&  0.0225  & - & No additional & 92.63\% \\
  \hline
  Chad & $B{32}$ & 4.529	  & 0.0016	& 0.597 & ITN campaigns & 91.42\%\\
  \hline
  The Congo  &  $B{21}$  &  4.3830 & 0.0068  & - & No additional& 25.15\%	 \\
  \hline
  Cote d'Ivoire & $B{31}$  &  4.7327 & 0.0041  &  0.5848	&  No additional& 54.004\% \\
  \hline
  DRC & $B{31}$  &  3.5274	& 0.0057 & 0.585 & No additional & $\ast$\\
  \hline
  Gui-Bissau & $B{31}$ &  1.5059 	& 0.129 & 0.591 & No additional & 6.89\%   \\
  \hline
  Nigeria & $B{21}$ & 4.777 & 0.013 & - & No additional & 118.01\% \\
  \hline
  Uganda & $B{31}$  &  3.8454	& 0.0195 & 0.598  & No additional  & 86.62\%\\
  \hline
  Angola & $B{32}$ & 1.9371 &	0.0185	& 0.691 & ITN campaigns \&  ITN efficacy & $\ast\ast$ \\
  \hline
  Burundi &  $B{31}$  & 1.6944 & 0.00038 & 0.593 & No additional & $\ast$ \\
  \hline
 Comoros & $B{31}$ & 0.4993  & 0.071 & 0.581 &No additional & 52.16\% \\
  \hline
  Guinea & $B{32}$	&  4.095 & 0.00022 & 0.787 & ITN campaigns \&  ITN efficacy &  $\ast$ \\
  \hline
  South Sudan &  $B{21}$ 	& 4.635 & 0.0471 & - & No additional & 151.50\% \\
  \hline
\end{tabular}}%
\label{tab:Table 5}
\vspace{0.5cm}
\caption{Optimal path and parameter values, targeted intervention, and relative increase of ITN usages (concerning 2022 as a baseline value) for achieving the 2030 milestone (Phase-II). $\ast$ indicates there is no change in ITN usages, and $\ast\ast$ indicates no change or even less coverage of ITN usage with higher bed net efficacy. Detailed discussion is in the main text.}
\end{center}

\end{table}

\begin{table}[H]
\begin{center}
\begin{tabular}{ |p{0.15\linewidth}|p{0.15\linewidth}|p{0.3\linewidth}|p{0.15\linewidth}|p{0.2\linewidth}|}
\hline
Cluster 1 & Cluster 2 & Cluster 3 & Cluster 4 & Cluster 5\\\hline
 \hline
Ethiopia, Gambia,  Mauritania,  Mozambique, Cote D'ivore,  Democratic Republic of Congo & Senegal,  Equatorial Guinea, Kenya,  Burkina Faso,  Sierra Leone, Tanzania, Nigeria &  Ghana, Togo, Malawi,  Somalia,  Cameroon, Central Africa Republic, Zimbabwe, Republic of Congo,  Guinea Bissau,  Uganda  & Gabon,  Zambia, Angola, Liberia, Burundi, Comoros &   Chad, Guinea, South Sudan  \\
\hline	   
\end{tabular}%
\vspace{1mm}
\caption{Countries classification in different clusters}
\label{tab:Table 6}
\end{center}
\end{table}

\begin{table}[H]
\begin{center}
\begin{tabular}{ |p{1.75cm}|p{2.25cm}|p{2.25cm}|p{3.25cm}|}
 \hline

 Cluster  &  Mean Biting rate ($\beta$) & Mean relative cost  ($r$) & Mean mosquito recruitment rate  ($\Lambda_v$) \\\hline
 \hline
   1 & 2.084 &  0.0068 & 0.2411 \\
   \hline
   2 & 1.069 &  0.2076 & 0.3643\\
   \hline
   3 & 1.630 & 0.0835  & 0.2796\\
   \hline
   4 &  0.954 & 0.0634 & 1.3044\\
   \hline
   5 &  5.837 & 0.1042 & 0.5484\\
    
  \hline	   

\end{tabular}%
\vspace{2mm}
\caption{Mean biting rate, relative cost and mosquito recruitment rate for different clusters} 
\label{tab:Table 7}
\end{center}
\end{table}

\begin{figure}[hbt!]
    \centering
  \includegraphics[width=1\linewidth]{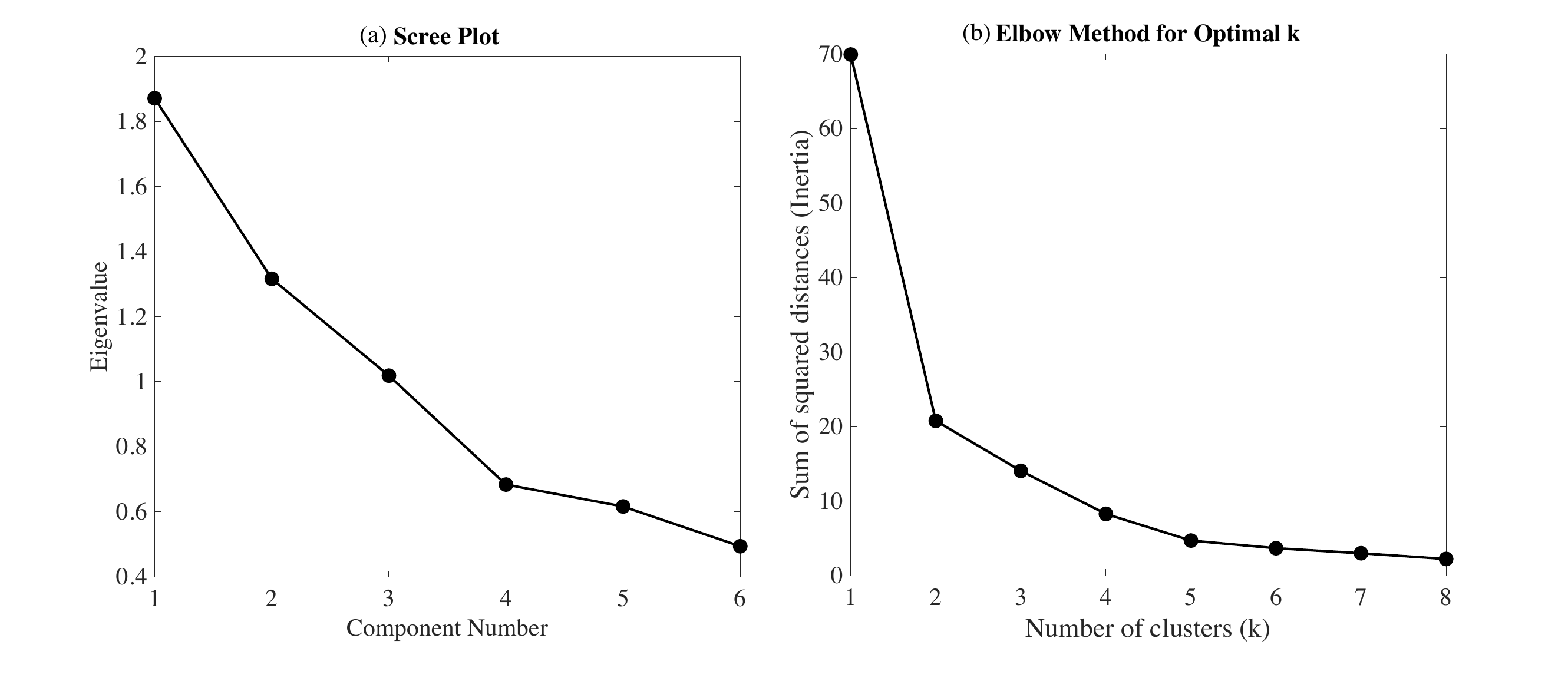}
   \caption{(a) Screen Plot showing components with Eigen values and (b) Plot of the inertia for different k}
    \label{fig:scree_elbow}
\end{figure}

\begin{figure}[hbt!]
    \centering
    \hspace{-1.6cm}
    \includegraphics[width=1\linewidth]{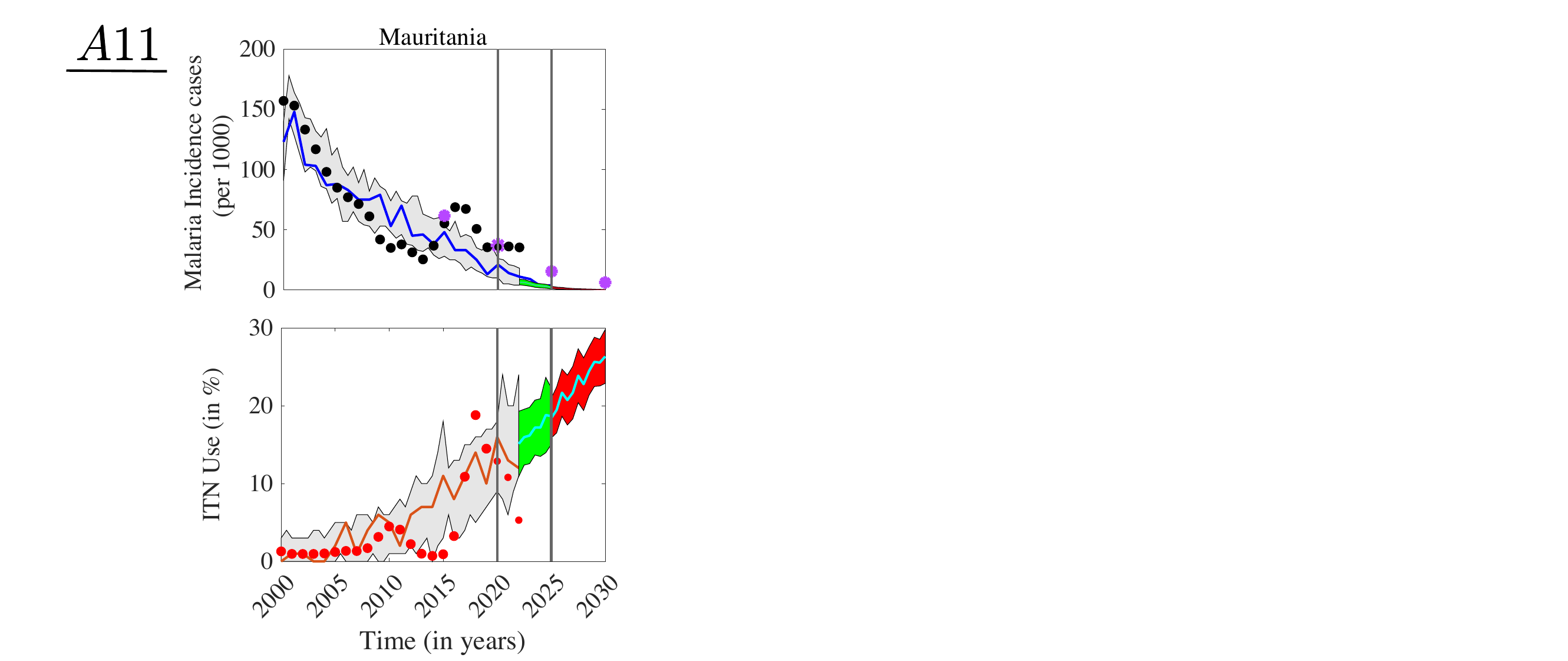}
     \vspace{-1mm} 
    \hspace{-2.7cm}
    \includegraphics[width=1\textwidth]{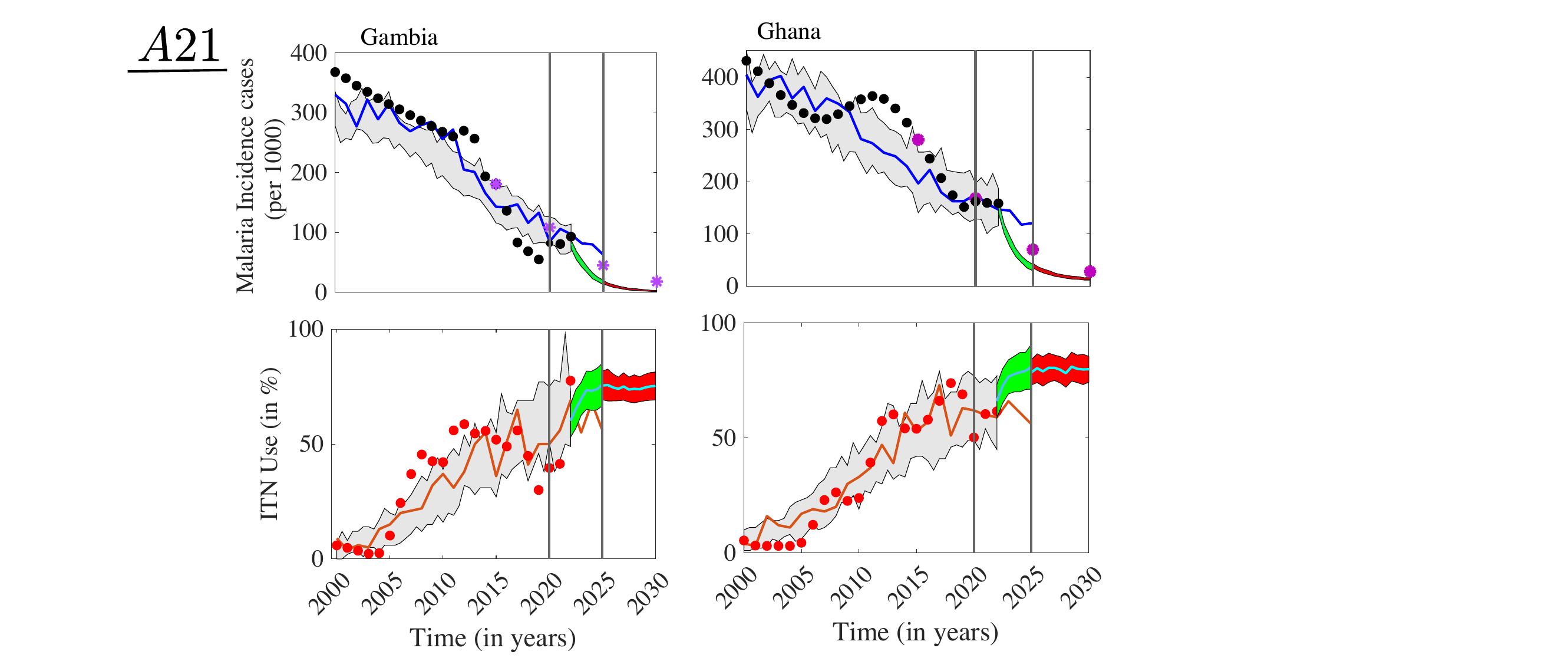}
   
     \vspace{1mm} 
     
     \caption{Countries follow path $A11$ and $A21$ for 2025 and 2030}
    \label{fig:path_A}
\end{figure}

\begin{figure}[hbt!]
    \centering
    \includegraphics[width=1\textwidth]{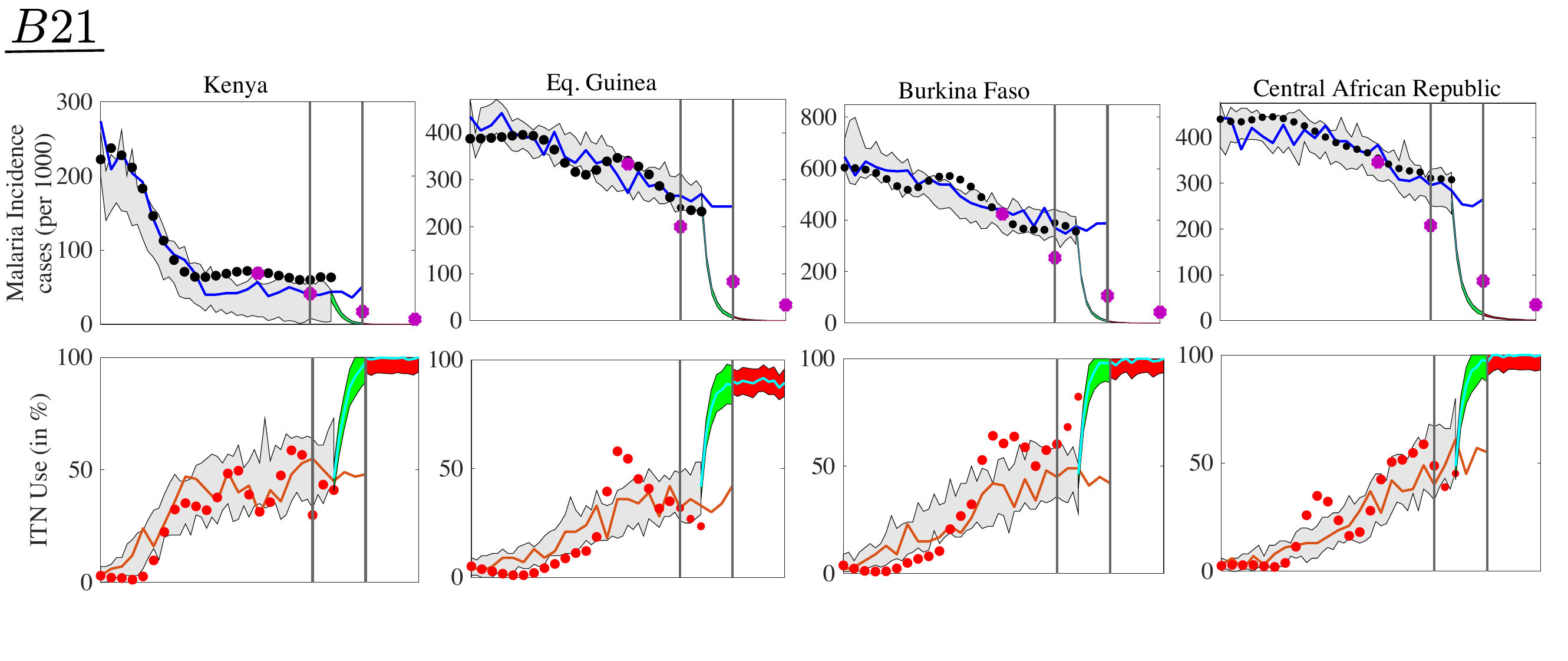}
   \vspace{-4mm} 
     \hspace{-0.1cm}
     
    \includegraphics[width=1\textwidth]{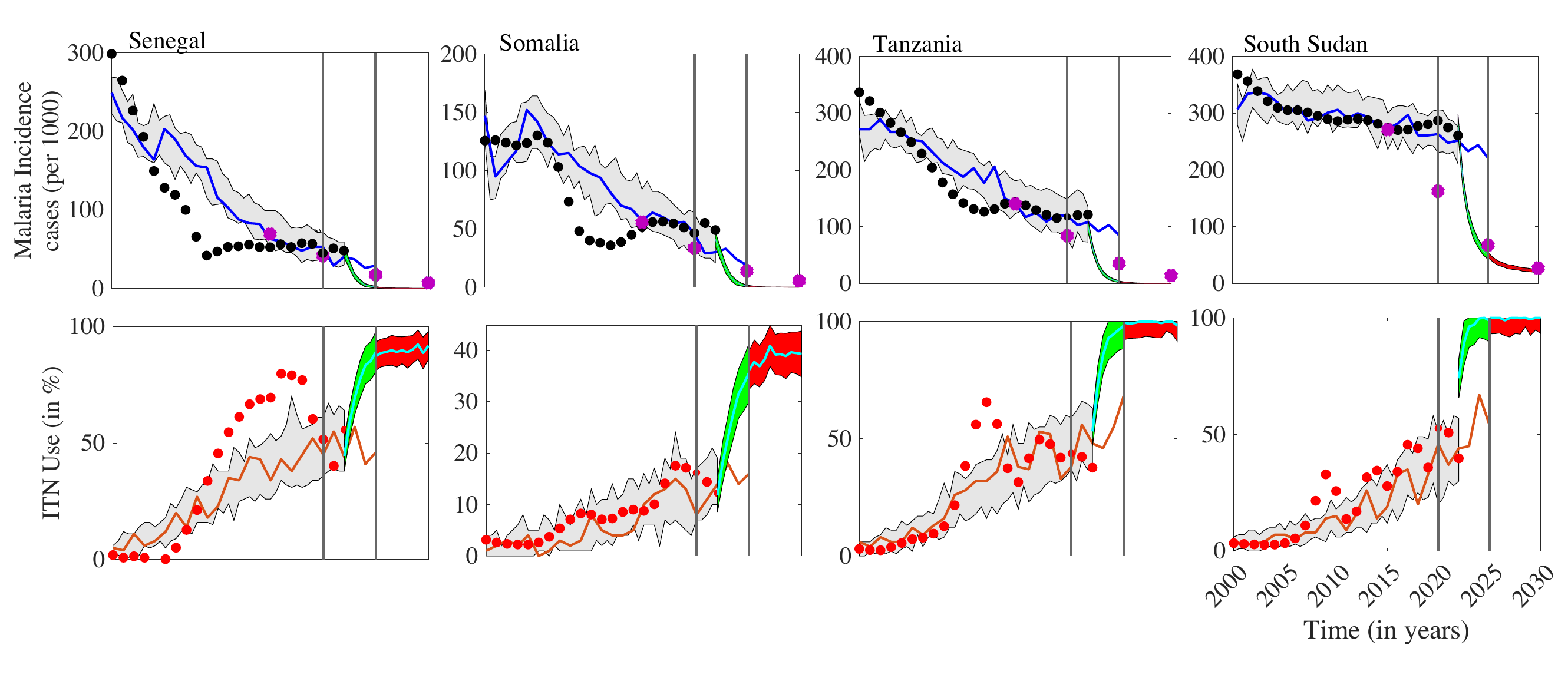}
    \vspace{-10mm} 

    \hspace{0.3cm}
     \includegraphics[width=1\textwidth]{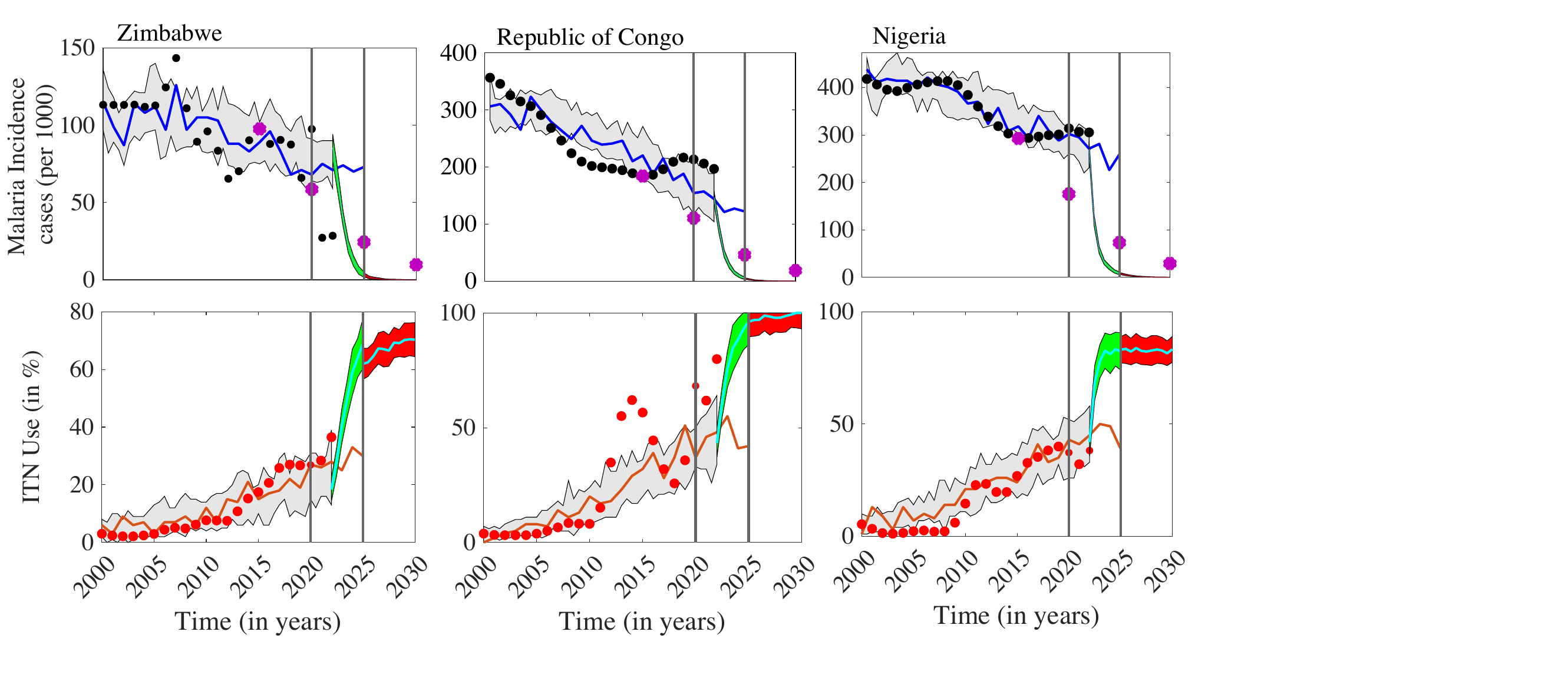}

    \caption{Countries follow $B{21}$  path for 2025 and 2030}
    \label{fig:B_21}
\end{figure}

\begin{figure}[hbt!]
    
     \hspace{-1.2cm}
      \vspace{8mm}
      \includegraphics[width=1\textwidth]{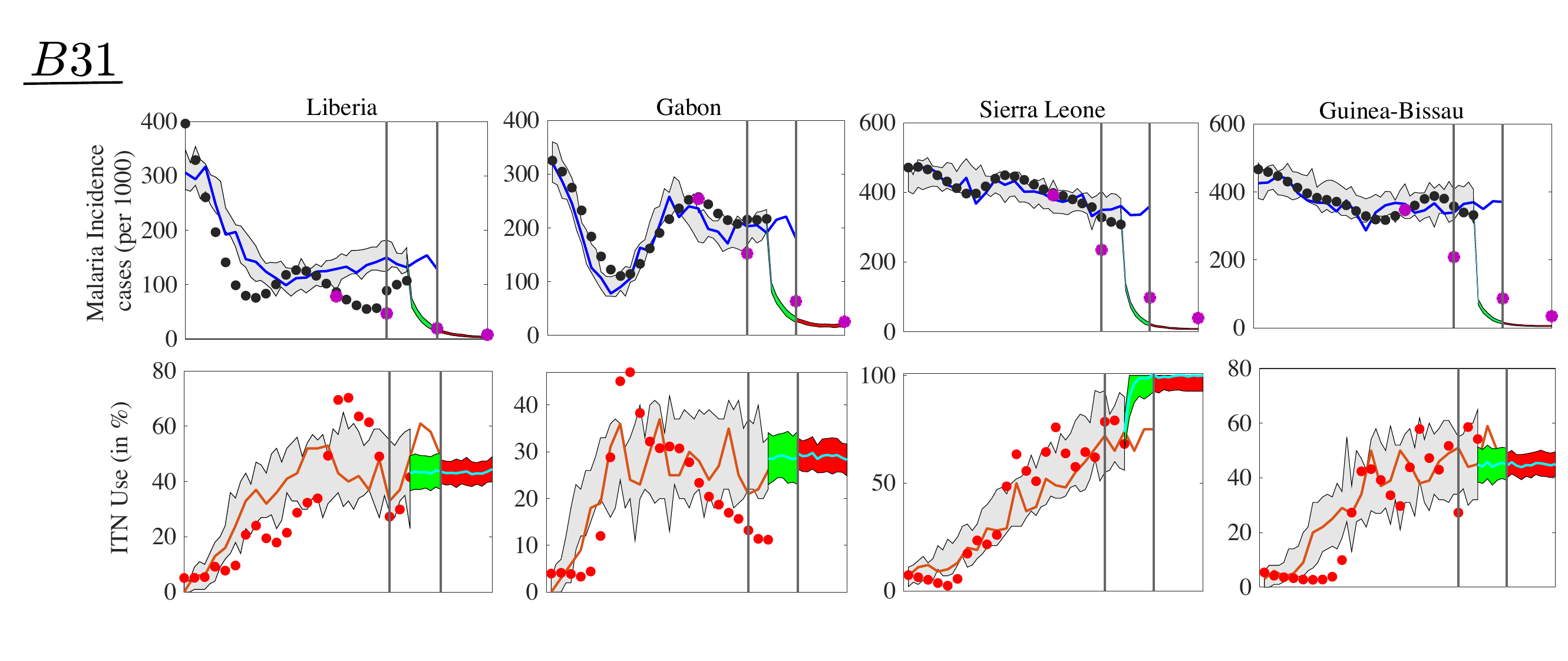}
    \hspace{-0.8cm}
    \vspace{-10mm} 
    
    \includegraphics[width=1\textwidth]{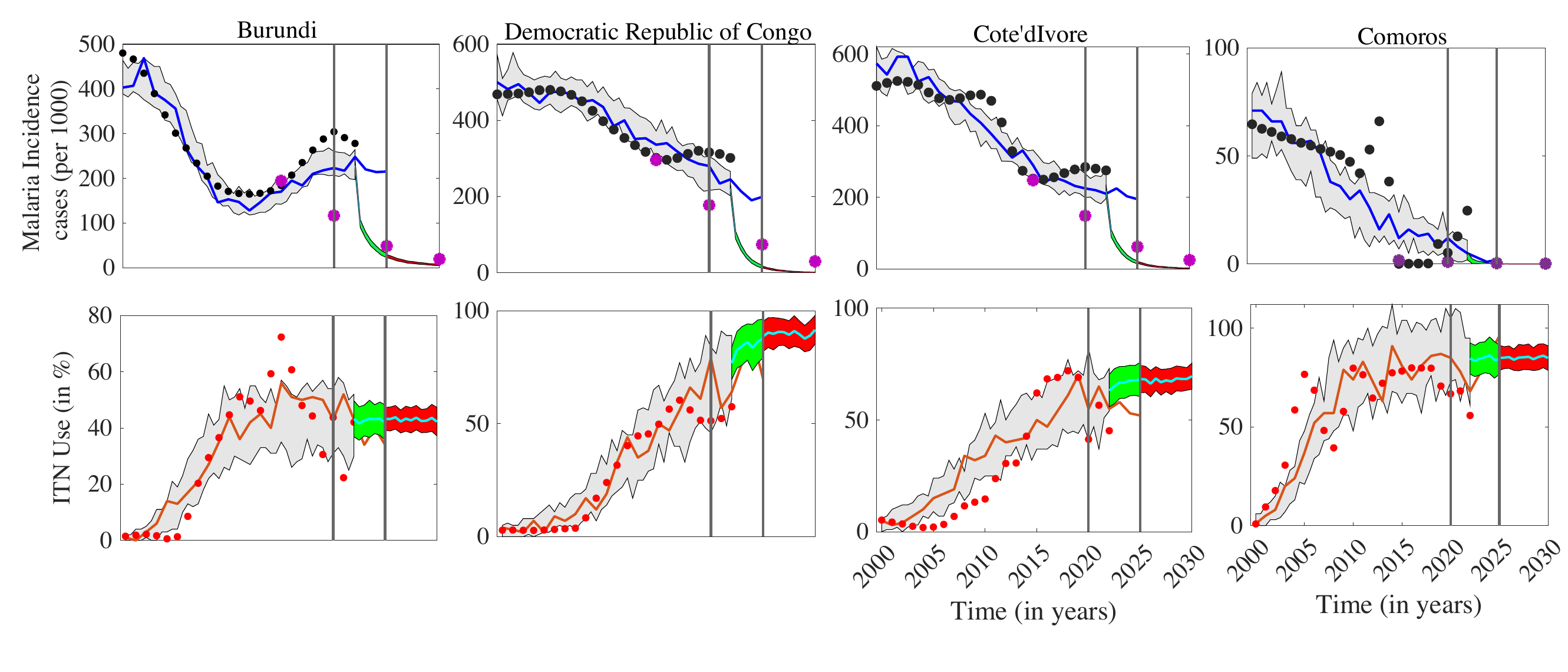}
     \vspace{-10mm} 
    
    \includegraphics[width=1\textwidth]{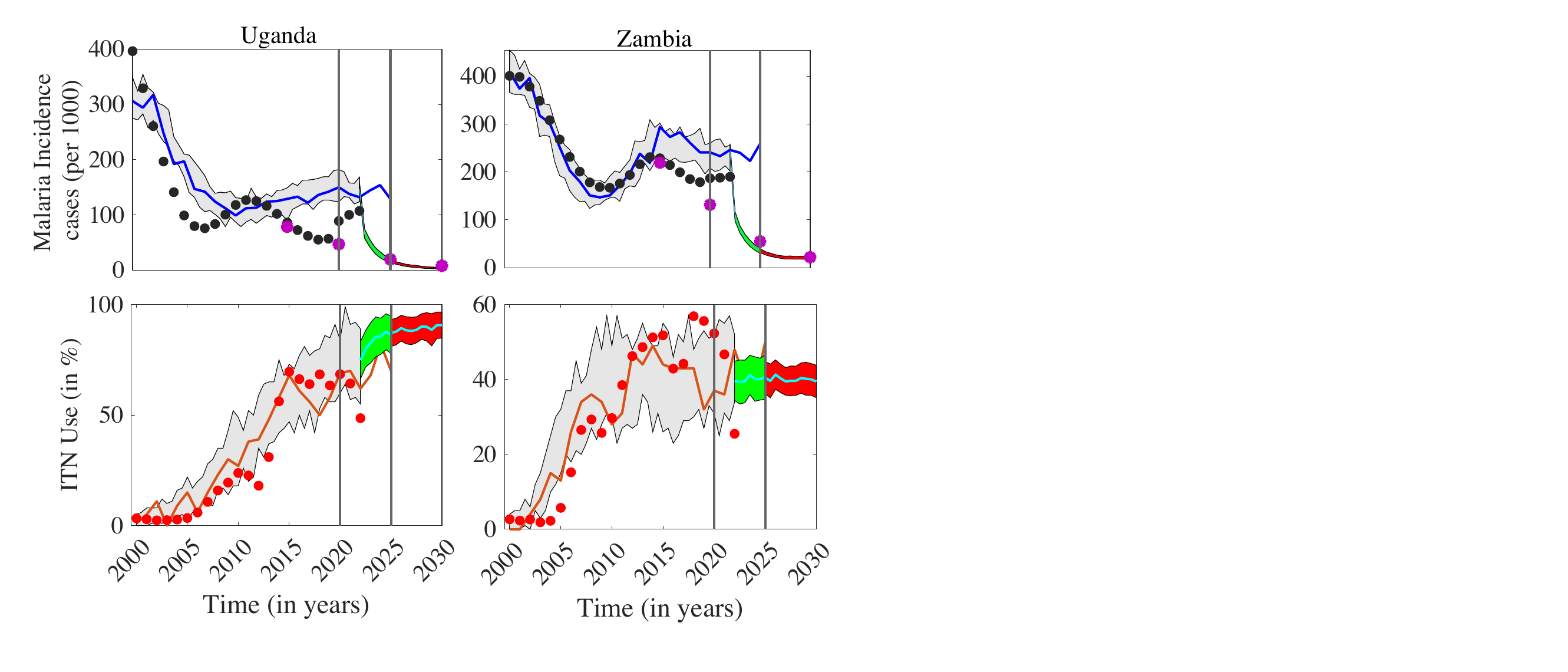}

    \hspace{2cm}
    \caption{Countries follow $B{31}$ path for 2025 and 2030}
    \label{fig:B_31}
\end{figure}

\begin{figure}[hbt!]
    
    \centering
     \includegraphics[width=1\textwidth]{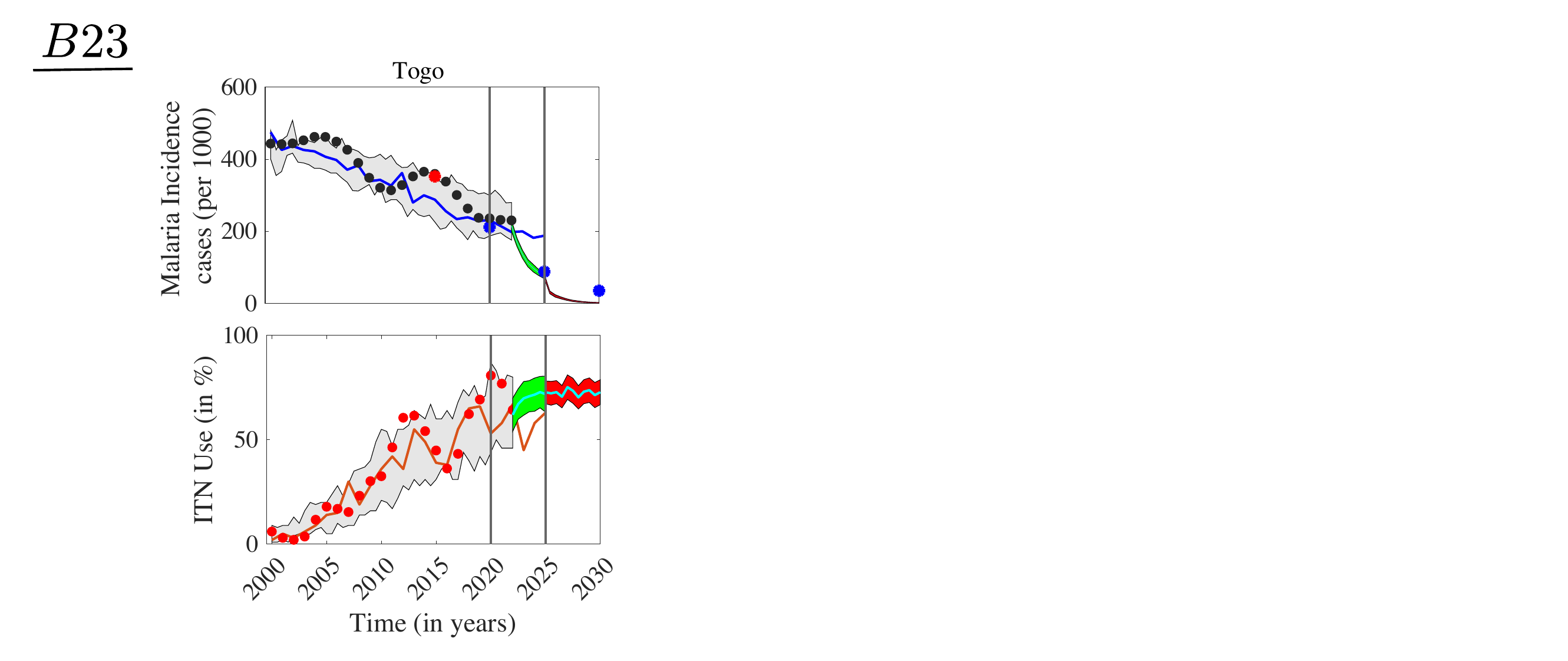}
      \hspace{1.8cm}
    \vspace{-10mm} 
\includegraphics[width=1\textwidth]{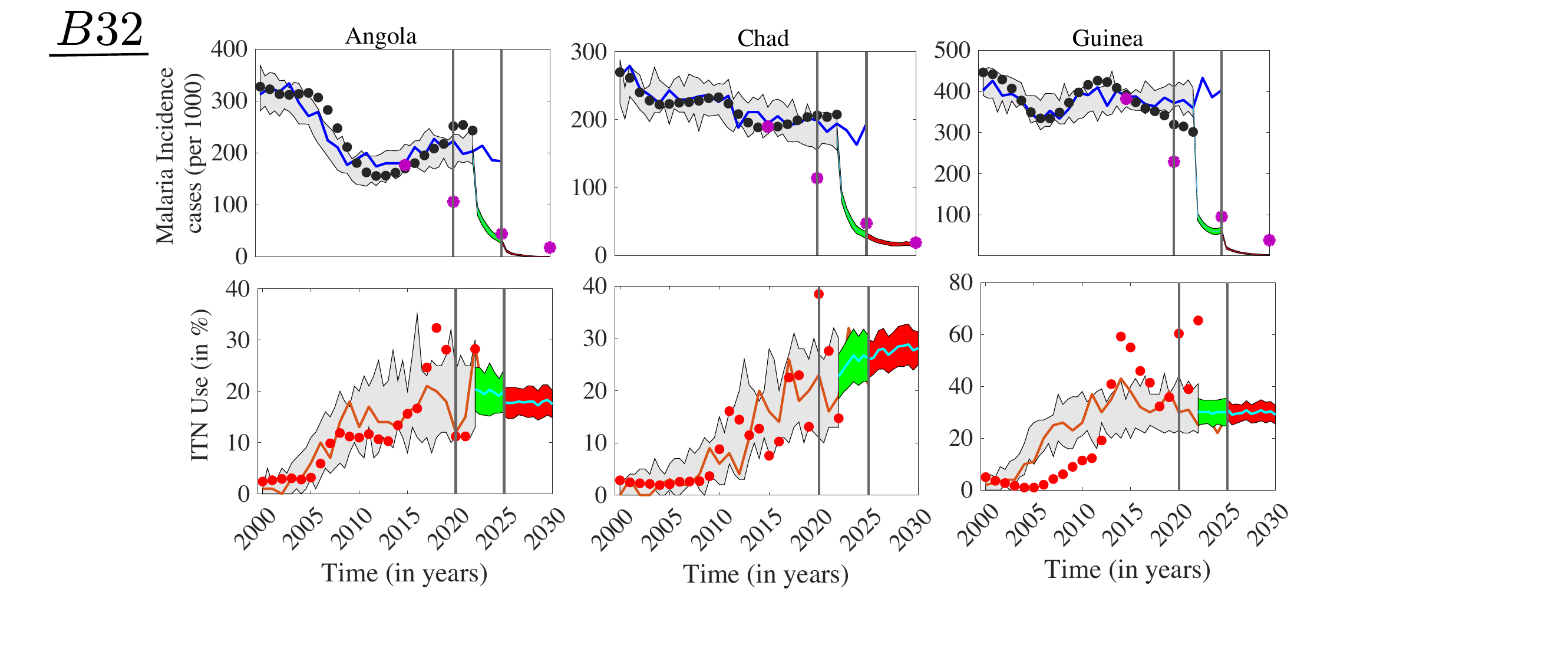}
\vspace{1cm}
 \caption{Countries follow $B{23}$ and $B{32}$ path for 2025 and 2030}
    \label{fig:B_23}
\end{figure}
\newpage

\bibliographystyle{abbrv}
\bibliography{main}